\documentclass[prd,twocolumn,showpacs,letterpaper,nofootinbib,superscriptaddress]{revtex4}
\usepackage{graphics}
\usepackage{epsfig}
\usepackage{amsmath}

\newcommand{\<}[1]{\langle {#1} \rangle}

\begin{document}

\title{Dark-Matter Electric and Magnetic Dipole Moments}

\author{Kris Sigurdson}
\email{ksigurds@tapir.caltech.edu}
\affiliation{California Institute of Technology, Mail Code 130-33, Pasadena, CA
91125}
\author{Michael Doran}
\affiliation{Department of Physics and Astronomy, Dartmouth College, 6127 Wilder Laboratory, Hanover, NH 03755}
\author{Andriy Kurylov}
\affiliation{California Institute of Technology, Mail Code 130-33, Pasadena, CA
91125}
\author{Robert R. Caldwell}
\affiliation{Department of Physics and Astronomy, Dartmouth College, 6127 Wilder Laboratory, Hanover, NH 03755}
\author{Marc Kamionkowski}
\affiliation{California Institute of Technology, Mail Code 130-33, Pasadena, CA
91125}


\begin{abstract}

We consider the consequences of a neutral dark-matter particle with a nonzero
electric and/or magnetic dipole moment. Theoretical constraints, as well as
constraints from direct searches, precision tests of the standard model, the
cosmic microwave background and matter power spectra, and cosmic gamma rays,
are included. We find that a relatively light particle with mass between an MeV
and a few GeV and an electric or magnetic dipole as large as $\sim 3\times
10^{-16}e$~cm (roughly $1.6\times10^{-5}\,\mu_B$) satisfies all experimental
and observational bounds.  Some of the remaining parameter space may be probed
with forthcoming more sensitive direct searches and with the Gamma-Ray Large
Area Space Telescope.

\end{abstract}



\maketitle

\section{Introduction} 

A wealth of observational evidence indicates the existence of considerably more
mass in galaxies and clusters of galaxies than we see in stars and gas.  The
source of the missing mass has been a problem since Zwicky's 1937 measurement
of the masses of extragalactic systems \cite{Zwicky:1937}.  Given the evidence
from galaxy clusters, galaxy dynamics and structure formation, big-bang
nucleosynthesis, and the cosmic microwave background that baryons can only
account for $\sim1/6$ of this matter, most of it must be nonbaryonic.  Although
neutrinos  provide the cosmological density of dark matter if their masses sum
to $\sim12$ eV, such particles cannot (essentially from the Pauli principle)
have a sufficiently high phase-space density to account for galactic
dark-matter halos \cite{Tremaine:we}; moreover, such masses are now
inconsistent with neutrino-mass measurements \cite{pdg}.  Theorists have thus
taken to considering for dark-matter candidates new physics beyond the standard
model. To date, the most promising candidates---those that appear in fairly
minimal extensions of the standard model and which coincidentally have a
cosmological density near the critical density---are a weakly-interacting
massive particle (WIMP), such as the neutralino, the supersymmetric partner of
the photon, $Z^0$ boson, and/or Higgs boson \cite{Jungman:1995df}, or the axion
\cite{axionreviews}. A considerable theoretical literature on the properties
and phenomenology of these particles has arisen, and there are considerable
ongoing experimental efforts to detect these particles.

In the absence of discovery of such particles, it may be well worth exploring
other possibilities.  Thus, an alternative line of investigation takes a more
model-independent approach and seeks to explore phenomenologically the possible
properties of a dark-matter particle.  Along these lines, for example,
constraints to strongly-interacting dark matter were considered in Ref.
\cite{Starkman:1990nj}; self-interacting dark matter has been considered
\cite{Spergel:1999mh}, and some have studied whether dark matter might be
charged \cite{Gould:gw} or have a millicharge
\cite{millichargeone,millichargetwo}.

Our investigation follows in spirit the latter possibility. In particular, dark
matter is so called because the coupling to photons is assumed to be
nonexistent or very weak, or else we would have presumably seen such particles
either through absorption or emission of radiation or in laboratory
experiments.  In this paper, we ask the question, ``How dark is `dark'?''  In
other words, how weak must the coupling of the dark-matter particle to the
photon be in order to be consistent with laboratory and astrophysical
constraints?  In the work on millicharged particles, a dark-matter coupling to
photons was assumed to arise from a tiny charge.

\begin{figure*}[htbp]
\resizebox{16cm}{!}{\includegraphics{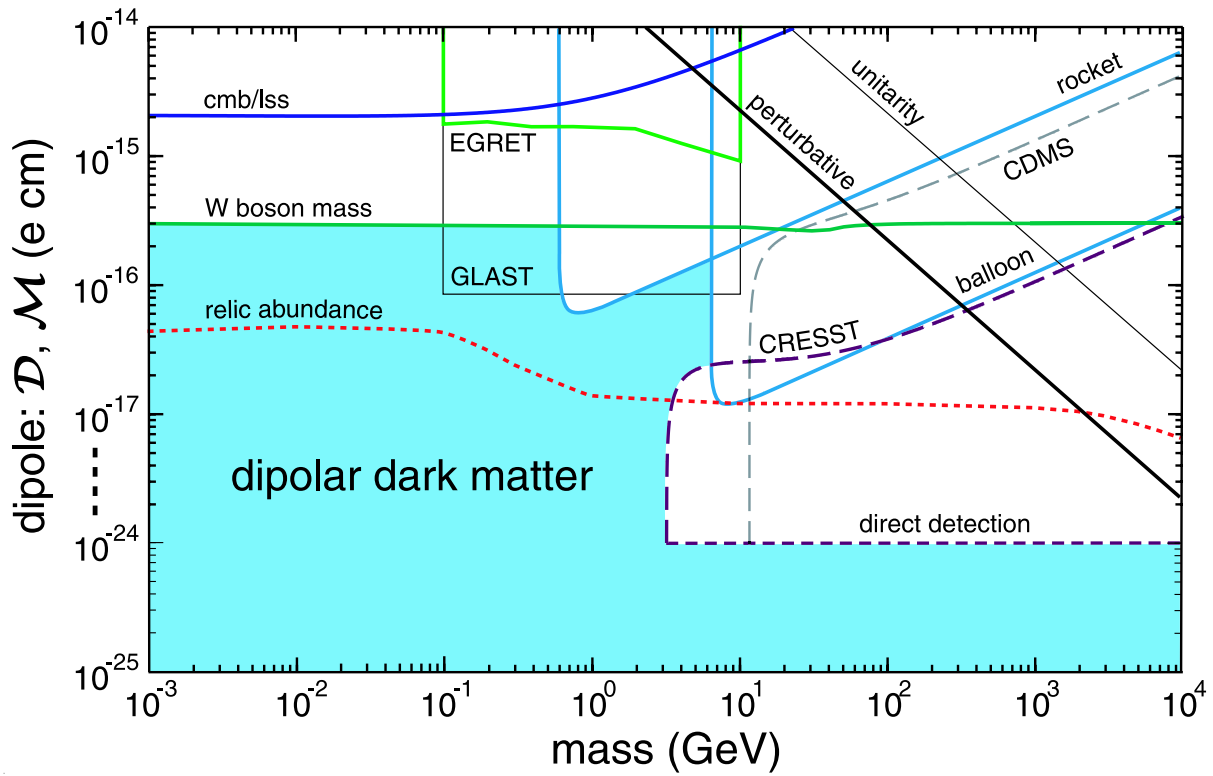}}
\caption{\label{fig:concord} The constraints on the dipolar-dark-matter
parameter space $[m_\chi,\,({\cal D,\,M})]$ that come from present-day searches
and experiments.  Viable candidates must lie in the shaded region, below the
solid lines and outside the long-dashed lines.  The short-dashed ``relic
abundance'' curve shows where the dark matter would have a cosmological density
$\Omega_\chi h^2=0.135$, assuming standard freezeout, no particle-antiparticle
asymmetry, and no interactions with standard-model particles apart from the
dipole coupling to photons.}
\end{figure*}

In this paper we consider the possibility that the dark matter possesses an
electric or magnetic dipole moment. The punch line, illustrated in
Fig.~\ref{fig:concord}, is that a Dirac particle with an electric or magnetic
dipole moment of order $\sim 10^{-17} e~$cm with a mass between an MeV and a
few GeV can provide the dark matter while satisfying all experimental and
observational constraints.\footnote{Throughout, we will quote numbers for both
the electric and magnetic dipole moments in units of $e$~cm, where $e$ is the
electron charge.  For reference, the Bohr magneton $\mu_B = e\hbar/2
m_e=1.93\times10^{-11}\,e$~cm in these units.  Also note that we work in
rationalized Gaussian units so that the fine-structure constant $\alpha\equiv
e^2/4\pi\hbar c \approx 1/137$, and in particle-physics units ($\hbar=c=1$)
$e^2 \approx 4\pi/137$ and $e \approx 0.303$.} 

In the following Section, we introduce the effective Lagrangian for the dipolar
dark matter (DDM) interaction with photons.  We discuss the relic abundance in
Section \ref{section3}.  Section  \ref{section4} presents constraints on
dark-matter dipole moments and masses that arise from direct searches at
low-background detectors as well as constraints from high-altitude
experiments.  Section  \ref{section5} discusses constraints due to precision
tests of the standard model, while Section  \ref{section6} discusses
constraints due to the cosmic microwave background and the growth of
large-scale structure.  We provide some concluding remarks in Section 
\ref{section7}.  An Appendix provides details of the calculation of the drag
force between the baryon and DDM fluids used in Section  \ref{section6}.

\section{Theory of Dipole Moments}
\label{sec:theory}

A particle with a permanent electric and/or magnetic dipole moment must have a
nonzero spin; we thus consider spin-1/2 particles.  Moreover, Majorana
particles cannot have permanent dipole moments, so we consider Dirac fermions. 
Since the spin and the magnetic dipole are both axial vectors, a magnetic
dipole moment can arise without violating any discrete symmetries.  However,
the electric dipole moment is a vector and thus requires time-reversal and
parity violation.

The effective Lagrangian for coupling of a Dirac fermion $\chi$ with a magnetic
dipole moment ${\cal M}$ and an electric dipole moment ${\cal D}$ to the
electromagnetic field $F^{\mu\nu}$ is
\begin{align}
     {\cal L}_{\gamma\chi} = 
          -\frac{i}{2}\bar{\chi}\sigma_{\mu\nu}({\cal M} 
          + \gamma_{5}{\cal D})\chi F^{\mu\nu}.
\label{Lint}
\end{align}
At energies low compared to the dark-matter mass, the photon is blind to
distinctions between ${\cal M}$ and ${\cal D}$ (unless time-reversal-violating
observables are considered). Hence, we can discuss limits to ${\cal D}$ which
equally apply to ${\cal M}$, except where noted.

On dimensional grounds, we expect the electric dipole moment to be ${\cal D}
\lesssim e \, m_\chi^{-1}\simeq 2\times 10^{-14}\, (m_p/m_{\chi})\,e$~cm, where
$m_p$ is the proton mass. Similar arguments also apply to the magnetic dipole
moment.  This limit is shown as the perturbative bound in
Fig.~\ref{fig:concord}, as violation of this bound would signal some
non-trivial or non-perturbative field configuration.  As we will see below,
more rigorous but slightly weaker upper limits can be set with unitarity
arguments.

These upper limits already simplify our analysis.  The phenomenology of charged
dark-matter particles is determined largely by the ability of these particles
to form atom-like bound states with electrons, protons, or each other. However,
dipolar dark matter cannot form such bound states. A neutral particle with a
magnetic moment will not form bound states with charged particles.  Curiously
enough, a neutral particle with an electric dipole moment (EDM) can indeed form
a bound state with an electron, as first noted by no less than Fermi and Teller
\cite{Fermi:1947}, but only if the dipole moment is greater than $0.639 ~ e \,
a_0=3.4\times10^{-9}\,e$~cm  (assuming $m_\chi\gg m_e$) where $a_0$ is the Bohr
radius.  For smaller values of the dipole, the electron ``sees'' both poles of 
the dipole and finds no stable orbit.  This critical electric dipole moment
scales inversely with the dipole-electron reduced mass, so a bound state with a
proton can occur if the dipole mass is $\gg m_p$ and ${\cal D} \gtrsim
1.8\times 10^{-12}e$~cm. As we will see below, such values for the EDM cannot
occur for a pointlike DDM.  Likewise, the weakness of the dipole-dipole
interaction prevents the formation of any stable dark-matter atoms.

The first cosmological constraint is that from big-bang nucleosynthesis (BBN). 
BBN requires that the effective number of relativistic degrees of freedom at
$T\sim$~MeV does not exceed the equivalent of roughly 0.2 neutrino species
\cite{bbn}.  Since the particles we are considering are Dirac particles, they
act like two effective neutrino species and thus cannot be relativistic and in
equilibrium at BBN.  Generally, such considerations rule out $m_\chi
\lesssim$~MeV, and so we restrict our attention in Fig. \ref{fig:concord} to
masses above an MeV.  Strictly speaking, if the dipole moment is $[{\cal
D},{\cal M}]\lesssim 10^{-22}\,e$~cm, and if the particle has no other
interactions with standard-model particles, then a particle of mass
$\lesssim$~MeV can decouple at a temperature $\gtrsim10$~GeV, and if so, it
will evade the BBN limit.

\section{Dark Matter Annihilation and Relic Abundance}
\label{section3}

\begin{figure}[b]
\begin{center}
\resizebox{5cm}{!}{\includegraphics{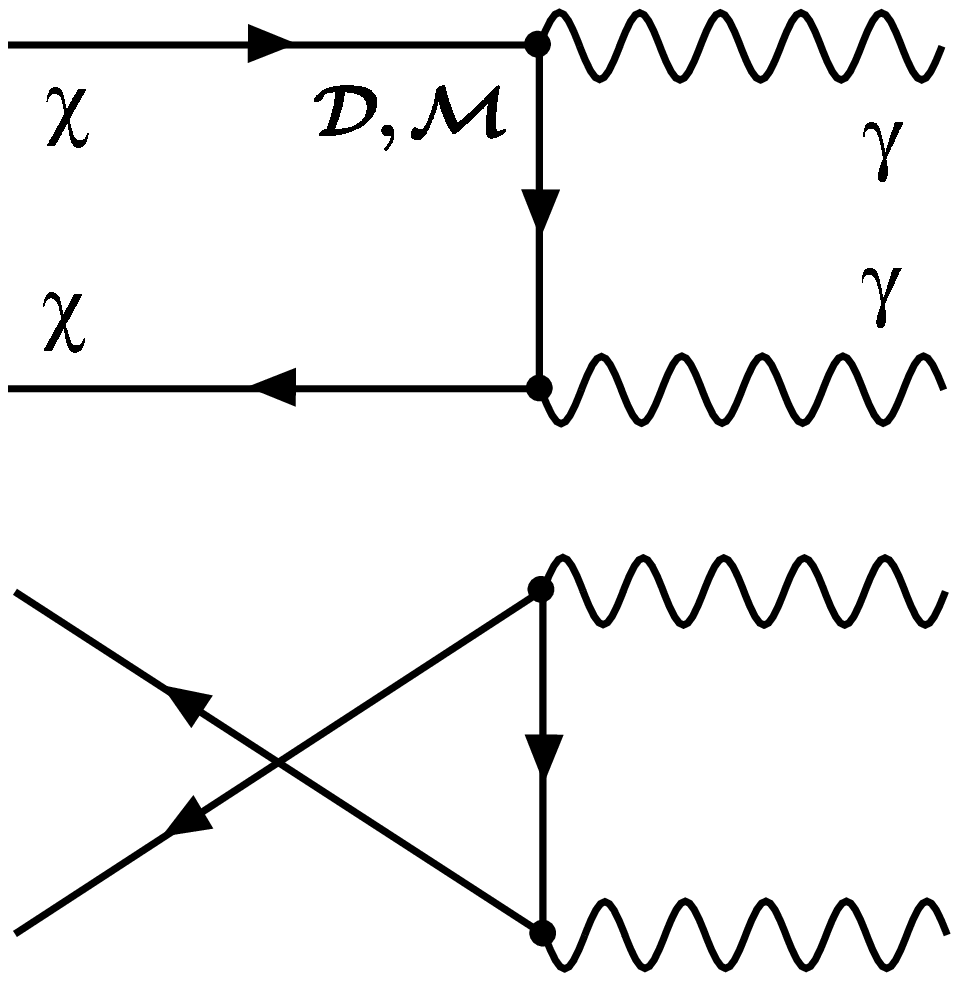}}
\caption{Feynman diagrams for annihilation of a DDM--anti-DDM
pair to two photons.}
\label{fig:ddtophotons}
\resizebox{4cm}{!}{\includegraphics{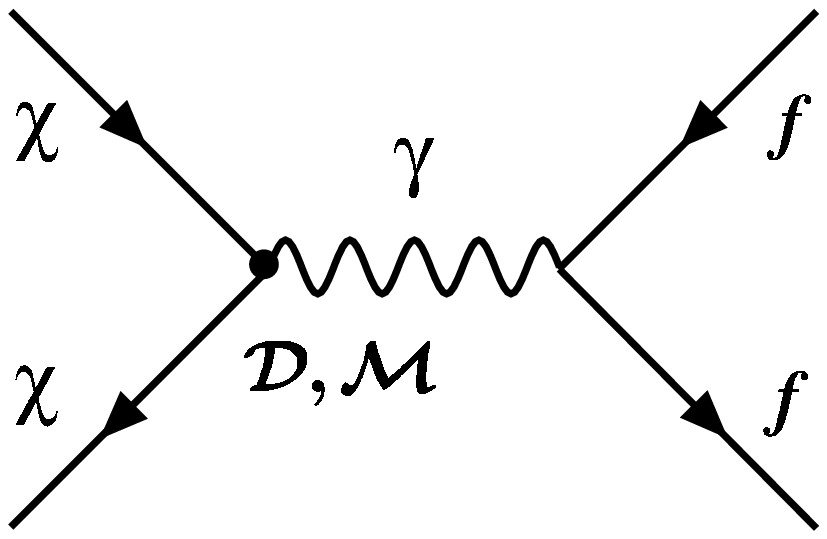}}
\caption{Feynman diagram for DDM--anti-DDM annihilation to
fermion-antifermion pairs.}
\label{fig:ddtoff}
\end{center}
\end{figure}

DDM particles can exist in thermal equilibrium in the early Universe when the
temperature $T \gg m_\chi$, and their interactions will freeze out when $T$
drops below $m_\chi$ resulting in some cosmological relic abundance. The mass
density of relic DDM particles is fixed by the cross section $\sigma_{\rm ann}$
for annihilation to all lighter particles times the relative velocity $v$
through (see, e.g., Eq. (5.47) in Ref. \cite{KolbTurnerbook}),
\begin{eqnarray}
    \Omega_\chi h^2 &\simeq& 3.8\times10^{7}\left({m_\chi \over
    m_p}\right){\ln\left(A/\sqrt{\ln  A}\right) / A} \cr
    &=&0.135(g_*/10)^{-1/2} \left({\sigma_{\rm ann} v \over
    5.3\times 10^{-26}\,{\rm cm}^3~{\rm sec}^{-1}} \right)^{-1}\cr
    & & \qquad \times {\ln \left[ A/\sqrt{\ln A}\right] \over 21},
\end{eqnarray}
where 
\begin{eqnarray}
     A &=& 0.038 \sqrt{g_*}m_{pl} m_\chi (\sigma_{\rm ann} v)\cr
      &=& {6.3\times10^9\,(g_*/10)^{1/2}\over (m_\chi/{\rm
      GeV})} \left({\sigma_{\rm  ann} v \over
    5.3\times 10^{-26}\,{\rm cm}^3~{\rm sec}^{-1}}
    \right),
\end{eqnarray}
assuming that annihilation takes place (as it does in our case) through the $s$
wave.  Here, $g_*$ is the effective number of relativistic degrees of freedom
at the temperature $T_f \sim m_\chi/A$ of freezeout. For the interaction of Eq.
(\ref{Lint}), DDM--anti-DDM pairs can annihilate to either two photons or to
charged particle-antiparticle pairs through the diagrams shown in Figs.
\ref{fig:ddtophotons} and \ref{fig:ddtoff}. The cross sections for these two
processes (to lowest order in $v$) are
\begin{eqnarray}
     \sigma_{\chi\bar\chi \to 2\gamma} v &=& ({\cal D}^4 + {\cal
     M}^4) m_\chi^2 / 2\pi\cr
       &=& 1.0 \times 10^{-33} m_{\rm GeV}^2 ({\cal D}_{17}^4+{\cal
       M}_{17}^4)\,{\rm cm}^3~{\rm sec}^{-1},
     \cr\cr
     \sigma_{\chi\bar\chi \to f\bar f} v &=& N_{\rm eff} \alpha ({\cal
     D}^2 + {\cal M}^2)\cr
       &=& 2\times10^{-27}\,N_{\rm eff}\,({\cal D}_{17}^2+{\cal
       M}_{17}^2)\,{\rm cm}^3~{\rm sec}^{-1}, 
\label{eqn:ann}
\end{eqnarray}
where $[{\cal D}_{17},\,{\cal M}_{17}]=[{\cal D,\, M}]/(10^{-17}~e~{\rm cm})$,
and $m_{\rm GeV}\equiv m_\chi/{\rm GeV}$. Here, $N_{\rm eff}=\sum_f Q_f^2
N_{cf}$ is the effective number of fermion-antifermion pairs with mass
$m_f<m_\chi$, $Q_f$ is the charge of fermion $f$, and $N_{cf}$ is the number of
color degrees of freedom for fermion $f$. ($N_{cf}=1$ for electrons.)  In the
standard model, annihilation can also occur to $W^+W^-$ pairs above threshold.
For $({\cal D}_{17},{\cal M}_{17}) (m_\chi/m_p) \lesssim 5000$,  fermions are
the dominant final state. The present-day mass density of DDM particles thus
depends primarily on the dipole moment. If such particles are to account for
the dark matter, then $\Omega_\chi h^2 = 0.135$ \cite{CMBconstraints}, and
$({\cal D}^2 +{\cal M}^2)^{1/2} \simeq 1.0 \times 10^{-17}\, e$~cm for  $m_\chi
\sim 1~$GeV.  The full mass dependence  of this result is shown in
Fig.~\ref{fig:concord}.

The cross sections in Eq. (\ref{eqn:ann}) are $s$-wave cross sections. 
According to partial-wave unitarity, the total $s$-wave annihilation cross
section must be $\sigma \lesssim 4\pi/m_\chi^2$ \cite{Griest:1989wd}, so that
$({\cal D,\, M}) m_\chi \lesssim 3$, fixed by the cross section for
annihilation to two photons. This limit is shown in Fig.~\ref{fig:concord}, as
is the more stringent, but less rigorous, limit $({\cal D,\, M}) \lesssim
e/m_\chi$.

Of course, the present-day mass density of DDM particles could differ from the
estimates obtained above.  To obtain these results, we assumed (1) that the
dipole interaction with photons is the only interaction of these particles; and
(2) that there is no particle-antiparticle asymmetry.  It is reasonable to
assume that in any realistic model, a dark-matter dipole interaction will arise
from loop diagrams involving other standard-model and new particles.  If so,
then there may be other contributions to the annihilation cross sections.  In
this case, the relic abundance will be smaller than we have estimated above. 
We thus conclude that if there is no particle-antiparticle asymmetry, our
estimates should be treated as an upper bound to the relic abundance, and the
$\Omega_\chi h^2$ curve in Fig. \ref{fig:concord} should thus be considered an
upper limit to the desired values of ${\cal D}$ and ${\cal M}$. On the other
hand, the relic abundance could also be increased if exotic processes increase
the expansion rate during freezeout \cite{KamTur90}.

If there is an asymmetry between $\chi$ and $\bar\chi$, then the relic
abundance is increased relative to our estimate.  In this case, however, the
present-day Universe should contain predominantly either particles or
antiparticles.  Although there is no {\it a priori} reason to expect there to
be a particle-antiparticle asymmetry, the observed baryon-antibaryon asymmetry
might lead us to expect an analogous dark-matter asymmetry, should the dark
matter be composed of Dirac particles.  It is possible such asymmetries have a
common origin.

Finally, we have assumed above that the particles freeze out when they are
nonrelativistic.  However, as the dipole moment is lowered for a given mass,
freezeout occurs earlier.  If the dipole moment is reduced beyond a certain
value, and if there are no other couplings to standard-model particles, then
freezeout will be relativistic.  These particles will then be roughly as
abundant as photons, and they will overclose the Universe by huge margins
unless their masses are $\lesssim$~few eV; even in this case they will violate
constraints to hot dark matter from the CMB and large-scale structure, and they
will also be unable, from the Tremaine-Gunn argument, to make up the dark
matter in Galactic halos.  The transition from nonrelativistic to relativistic
freezeout occurs (again, assuming no non-dipole interactions with
standard-model particles) for $m_\chi{\cal D}_{17}^2 \lesssim10^{-10}$~GeV for
$m_\chi\gtrsim$~MeV, and for $m_\chi\lesssim$~MeV, at $m_\chi{\cal
D}_{17}^{4/3} \lesssim200$~MeV.

\section{Direct Detection}
\label{section4}

\begin{figure}[b]
\begin{center}
\resizebox{4cm}{!}{\includegraphics{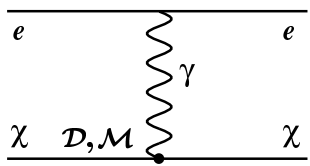}}
\caption{Feynman diagrams for elastic scattering of an electron
from a DDM particle.}
\label{fig:dn}
\end{center}
\end{figure}

The diagram for scattering of a DDM particle with a particle of charge $Ze$
occurs through the exchange of a photon, as shown in Fig. \ref{fig:dn} (not
unlike the electron-neutron interaction \cite{Foldy:1951}).  In the
nonrelativistic limit, the differential cross section for this process is given
by,
\begin{align}
     \frac{d \sigma}{d \Omega} = 
     \frac{Z^2 e^2 \left({\cal D}^2+{\cal M}^2\right)}{8\pi^2 v^2
     (1 - \cos\theta)},
\label{eqn:dn}
\end{align}
where $v$ is the relative velocity. Through this interaction the DDM might
produce a signal in a direct-detection experiment.  Although the total cross
section is formally infinite, the divergence comes from the
small-momentum-transfer scatterings that will be screened by atomic electrons. 
Roughly speaking, then, the DDM-nucleus elastic-scattering cross section will
be $\sigma\sim (Ze)^2 ({\cal D}^2+{\cal M}^2)/2\pi v^2 \simeq 6.4 \times
10^{-32}\,Z^2 ({\cal D}_{17}^2+{\cal M}_{17}^2)$~cm$^2$, using
$v\sim10^{-3}\,c$.  Current null searches in germanium detectors
[$(A,Z)=(76,32)$] correspond for masses $m_\chi\sim10$~GeV to a rough upper
limit to the cross section $\sim10^{-42}$~cm$^2$ \cite{CDMS04}, thus ruling out
any dipole moment $({\cal D}_{17}^2+{\cal M}_{17}^2)^{1/2}\gtrsim 10^{-7}$. 
This is shown in Fig. \ref{fig:concord} as the horizontal dashed line at ${\cal
D}=10^{-24}\,e$~cm.  Note that the cross-section limit depends (and increases)
with mass at higher masses; the curve appears as a horizontal line simply
because of the break in scale on the $y$ axis.

This seems like a very stringent limit, especially considering the value,
${\cal D}_{17}\sim1$, favored for the correct cosmological density.  However,
if the dipole moment becomes large enough, the particles will be slowed in the
rock above the detector and thus evade detection in these underground
experiments.  In order to determine the magnitude of the dipole moment for this
to occur, we need to calculate the stopping power $dE/dx$ for the particle as
it passes through the atmosphere and then the rock. Since elastic scattering
takes place through exchange of a photon, it leads to a long-range interaction
and, as we have seen above, a formally divergent elastic-scattering cross
section.  The calculation of the stopping power thus parallels that for
ionization loss due to Coulomb collisions, with two important differences.
First of all, since the long-range force is $\propto r^{-3}$, as opposed to
$r^{-2}$ for Rutherford scattering, stopping occurs via scattering from nuclei,
rather than electrons.  Second, since this interaction falls more rapidly with
radius than the Coulomb interaction, the stopping power is due primarily to
hard scatters at small impact parameter, rather than soft scatters at a wide
range of impact parameters. 

Our result for the stopping power due to scattering from nuclei of charge $Z$
is
\begin{equation}
     {dE\over dx} = -n_N \int T{d \sigma } 
     = -n_N {(Ze)^2 ({\cal D}^2+{\cal M}^2) \mu^2 c^2
     \over 2 \pi m_N},
\label{eqn:dEdx}
\end{equation}
where the kinetic-energy transfer in a single collision is $T = p^2(1 -
\cos\theta)/m_N$, $x$ is the depth, and $\mu=\mu[m_\chi,m_N]=m_\chi
m_N(m_\chi+m_N)^{-1}$ is the reduced mass. For very weak WIMP interactions with
nuclei, the most restrictive limits on the WIMP-proton cross section (the
smallest upper bounds) are obtained from null searches from experiments that
are deepest underground (so as to reduce the background).  However, the most
restrictive constraints on the cross section at the upper end of the excluded
range of cross sections will come from the shallowest underground experiment
with a null result.  From Eq. (\ref{eqn:dEdx}), we find that the dark-matter
particles will only penetrate to a depth $x = (E_{i} - E_{f})/|dE/dx|$ where
$E_i = \frac{1}{2} m_\chi v^2$ is the initial dark-matter kinetic energy and
$E_f$ is the final energy. For a stopped particle, $E_f=0$. However, the
particle only needs to lose enough energy for $E_f$ to drop below the detection
threshold for a particular experiment. Equating the maximum kinetic energy
transferable in a collision ($\theta = \pi$) to the threshold detectable
nuclear-recoil energy ($E_{\rm th}$), we find the velocity must be slowed to
$v_{f}^2 = m_d E_{\rm th} / 2 \mu[m_\chi,m_d]^2$, where $m_d$ is the mass of
the nuclei in the detector, and $\mu[m_\chi,m_d]$ is the DDM-nucleus reduced
mass. Hence, the final dark-matter energy must be $E_f = m_\chi m_d E_{\rm
th}/4 \mu[m_\chi,m_d]^2$. 

So far we have assumed that the particle loses energy but is not significantly
deflected in each scatter; this will be a good approximation if $m_\chi\gg
m_N$.  However, when $m_\chi\lesssim m_{\rm Si}\simeq 26$ GeV, the dark-matter
particle may be backscattered upon encountering a terrestrial nucleus, rather
than simply being slowed without deflection.  In this case, the particle will
diffuse, undergoing $N\sim m_N/2m_\chi$ scatters before coming to rest.  If so,
the penetration depth will be reduced by an additional factor of $\sim
N^{-1/2}$.  We thus replace $dE/dx \rightarrow dE/dx [1+(m_N/2m_\chi)^{1/2}]$
in our expression for the penetration depth.

Then, for a given shielding thickness $L$, in meters water equivalent (mwe), we
invert the expression for the stopping distance to obtain the following bound
on the dipole strength:
\begin{equation}
{\cal D}^2+{\cal M}^2 > {\frac{1}{2} m_\chi v^2 - 
\frac{1}{4} {m_\chi m_d \over \mu[m_\chi,m_d]^2} E_{\rm th} \over
{e^2 \over 2 \pi} L \sum_i f_i Z_i^2 
{\mu[m_\chi, m_i]^2 \over  m_i^2} [ 1 +(m_i/2 m_\chi)^{1/2}]}
\label{Dscatteringbound}
\end{equation}
where the index $i$ sums over the composition of the shielding material, and
$f_i$ is the fractional composition by weight. We use a realistic model of the
composition of the Earth (chemical composition by weight:  O [46.6\%], Si
[27.7\%], Al [8.1\%], Fe [5\%], Ca [3.6\%], Na [2.8\%], K [2.6\%], Mg [2.1\%])
and atmosphere ($10$ mwe consisting of a 4:1 ratio of nitrogen to oxygen),
although the resulting bounds do not change substantially if we ignore the
atmosphere and approximate the Earth's crust as entirely composed of Si. We
take the initial DDM velocity to be 300 km~sec$^{-1}$.

The shallowest underground experiment with a strong null result is the Stanford
Underground Facility (SUF) run of the Cryogenic Dark Matter Search (CDMS)
\cite{Akerib:2003px}, which was situated at a depth of 16 mwe. With a detector
energy threshold of $E_{\rm th}=5$~keV, it is sensitive to DDM masses down to
$m_\chi \sim 10$~GeV. Near this threshold we find that DDM particles are
stopped by the shielding for ${\cal D}_{17} \gtrsim 20$. This bound grows more
prohibitive with increasing mass, as illustrated in Fig.~\ref{fig:concord}. The
Cryogenic Rare Event Search with Superconducting Thermometers (CRESST) 
\cite{cresst}, though at a depth of 3800 mwe, extends to slightly lower masses,
having a detector threshold energy $E_{\rm th}=0.6$~keV. Near $m_\chi\sim
1~$GeV the minimum dipole strength is ${\cal D}_{17} \gtrsim 2$. However, there
are no limits from underground experiments for DDM masses below 1~GeV.

Two airborne experiments---unobscured by the atmosphere or rock---which have
closed the windows on some forms of strongly-interacting dark matter
\cite{Starkman:1990nj,McGuire:2001qj}, also restrict dark-matter dipole
moments. To determine the predicted signal at a detector, we recast Eq.
(\ref{eqn:dn}) as the cross section per energy transfer, whereby ${d \sigma / d
T} = {Z^2 e^2 ({\cal D}^2 + {\cal M}^2) / 4 \pi v^2 T}$. The event rate (per
time, energy, and unit mass of detector) is
\begin{eqnarray}
R &=& N_N (0.3 ~{\rm GeV}\,{\rm cm}^{-3}) {v\over m_\chi}
{d\sigma\over dT} \cr
&=& 2.3 \,({\cal D}_{17}^2 + {\cal M}^2) {m_p \over m_\chi} \left({{\rm keV} \over
T}\right) {\rm sec}^{-1}\,{\rm keV}^{-1}\,{\rm g}^{-1},
\end{eqnarray}
where $N_N$ is the number of nuclei per gram of material. 

The silicon semiconductor detector flown on a balloon in the upper atmosphere
by Rich {\it et al.} \cite{Rich:1987st} observed an event rate of $\sim 0.5\,
{\rm counts}\, {\rm sec}^{-1}\, {\rm keV}^{-1}\,{\rm g}^{-1}$ nuclear recoils
in the lowest energy bin at $2~$keV.  For dark-matter masses above the
threshold $\sim 7~$GeV, we thus require $({\cal D}_{17}^2+{\cal M}_{17}^2)
(m_p/m_\chi) < 0.2$.  

The X-ray Quantum Calorimeter (XQC) detector flown on a rocket by McCammon {\it
et al.} \cite{McCammon:2002gb}, was designed to probe the soft x-ray
background. However, it serendipitously provides a tight constraint on dipolar
dark matter. To predict the expected number of events, we start by computing
the number of DDM particles that could impinge on the detector: $N_\chi =
n_\chi v A t = 3\times 10^7 m_p/m_\chi$, where $n_\chi$ is the galactic number
density of dark-matter particles, $v$ is their velocity, the cross-sectional
area of the XQC detector is $A=0.33~$cm${}^2$, and the rocket flight was about
$t\sim 120~$seconds. The chief property of the XQC detector is the $14~\mu$m
thick Si substrate above the thermistor, providing a target of $N_N\sim
6.5\times 10^{19}~$nuclei/cm${}^2$. Thus, the event/energy count $N_N N_\chi
d\sigma/dT$ integrated over the 25--100 eV energy bin gives a predicted $\sim
0.38 \,({\cal D}_{17}^2+{\cal M}_{17}^2) (m_p/m_\chi)$ events compared to the
$\sim 10$ observed events. Since the detector has a $25 ~$eV threshold, energy
transfer by dark-matter particles as light as $\sim 1~$GeV can be detected.
Altogether, the balloon and rocket experiments exclude a wide range of dipole
strengths and masses, as illustrated in Fig.~\ref{fig:concord}.

\section{Constraints from Precision Measurements}
\label{section5}

We now consider the limits placed on DDM due to precision tests of the standard
model. Our use of perturbation theory is valid provided the energy scale of the
interaction ${\cal E}$ satisfies $({\cal D,\,M}) {\cal E} \ll 1$. In addition,
we require that the DDM mass satisfies $({\cal D,\,M}) m_\chi \lesssim 1$,
equivalent to the unitary bound \cite{Griest:1989wd}, which ensures the
self-consistency of the local operator in Eq.~(\ref{Lint}). Indeed, if
$\Lambda$ is the energy scale at which a dipole is generated then one generally
expects $\left({\cal D,\,M}\right)\Lambda \sim 1$. In ${\cal L}_{\gamma\chi}$
we assume that all interacting particles with masses greater than $\Lambda$
have been integrated out. Consequently, one must have at least $m_\chi <
\Lambda$ for the dark matter to be dynamical, which also yields $({\cal D,\,M})
m_\chi \lesssim 1$.

\subsection{Muon Anomalous Magnetic Moment}

\begin{figure}[t]
\begin{center}
\resizebox{4cm}{!}{\includegraphics{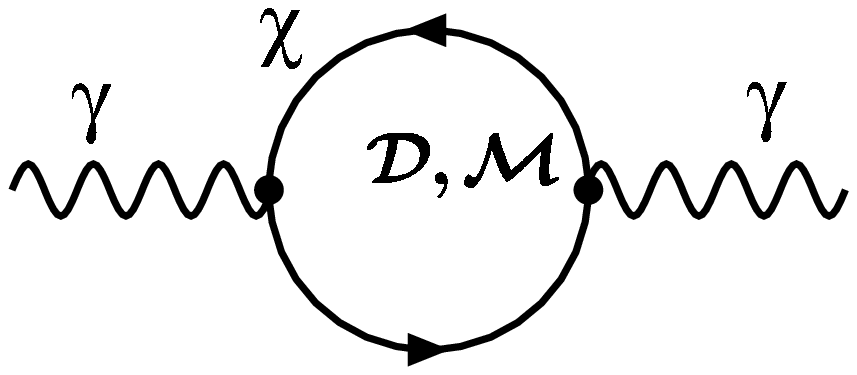}}
\caption{One-loop correction to the photon self-energy induced by
dipole moments $\cal{M},\cal{D}$ of the dark-matter particle.}
\label{fig:photon-prop}
\resizebox{3cm}{!}{\includegraphics{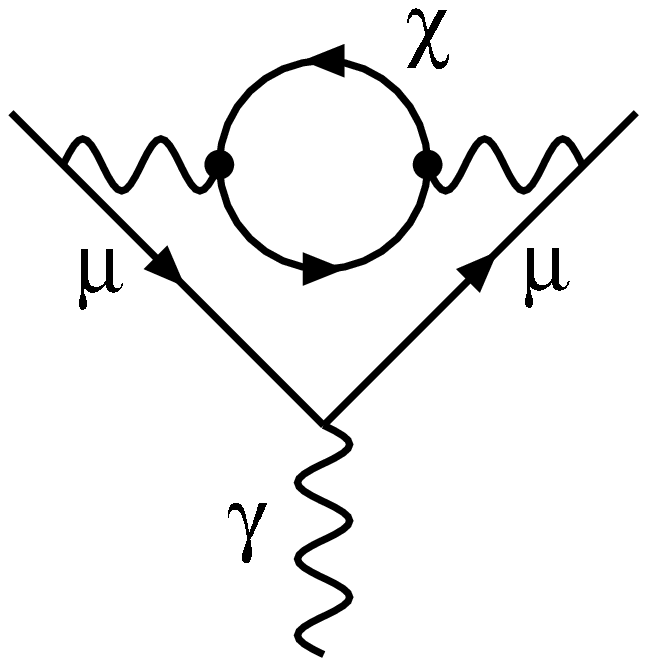}}
\caption{Lowest-order correction to the muon anomalous magnetic
moment induced by dipole moments of the dark-matter particle.}
\label{fig:mug2}
\end{center}
\end{figure}

The interaction in Eq.~(\ref{Lint}) contributes to the photon propagator via
the diagram shown in Fig.~\ref{fig:photon-prop}. The photon-DDM interaction
vertices are either both electric or magnetic dipolar; the mixed diagram where
one vertex is magnetic and the other is electric is proportional to
$\epsilon_{\mu\nu\rho\lambda}F^{\mu\nu}F^{\rho\lambda}=0$ for photons with
equal momenta. The sum of the diagrams produces the following contribution to
the photon vacuum-polarization tensor:
\begin{eqnarray}
\Pi^{\mu\nu}(q^2) &= &\Pi(q^2) \left(q^2 g^{\mu\nu}-
q^\mu q^\nu\right) \approx \beta q^2 \left(q^2 g^{\mu\nu}-
q^\mu q^\nu\right),\cr \beta &=& {{\cal D}^2 + {\cal M}^2\over 8
\pi^2}\left(1 - \frac{1}{3}\ln{m_\chi^2 \over \mu^2}\right),
\end{eqnarray}
where the photon momentum is taken to be small, $q^2 \ll m_\chi^2$ (resulting
in $\beta q^2 \ll 1$), and $\mu$ is the renormalization scale, which should be
smaller than $\Lambda$. We take $\mu\lesssim 1$ TeV for our estimates. With
this self-energy correction, the photon propagator for $\beta q^2\ll 1$ can be
written as
\begin{equation}
-i g^{\mu\nu} \left({1\over q^2} - {1 \over q^2 - 1/\beta}\right).
\end{equation}
The second term above generates a correction to the muon gyromagnetic ratio
$\delta a_\mu = -\alpha m_\mu^2 \beta/3 \pi$. Interestingly, this contribution
is not explicitly suppressed by the DDM mass. In view of recent measurements
\cite{Bennett:2002jb} and comparison with the SM predictions, we require that
$\delta a_\mu$ does not exceed $10^{-9}$, whereby $({\cal D}^2+{\cal
M}^2)^{1/2} <6\times 10^{-15}~e$~cm. The order of magnitude of this result can
be obtained on dimensional grounds, if we consider that the DDM dipole moment
contributes to $a_\mu$ via at least a two-loop graph (see Fig.~\ref{fig:mug2}),
with two electromagnetic couplings and two dipole couplings. Including a factor
$1/16\pi^2$ per loop, one obtains the estimate,
\begin{equation}
\delta a_\mu \sim {e^2 \over (16 \pi^2)^2} \left({\cal D}^2 +
{\cal M}^2\right){\cal E}^2,
\end{equation}
where ${\cal E}$ is the characteristic energy scale for the process. In the
case of the muon, ${\cal E}\sim m_\mu$, which reproduces the rigorous result to
within an order of magnitude.

\subsection{Electric Dipole Moments}

\begin{figure}[t]
\begin{center}
\resizebox{4cm}{!}{\includegraphics{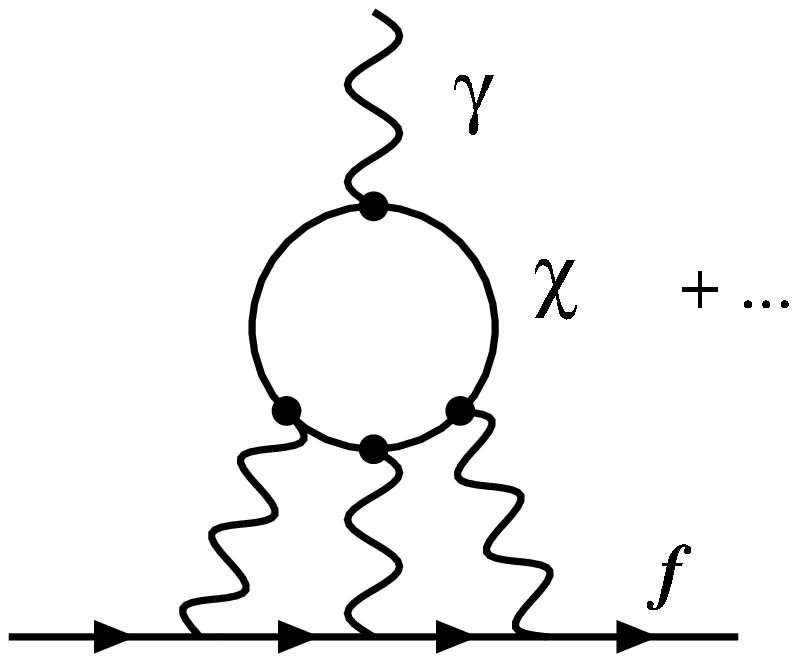}}
\caption{Three-loop contributions to the EDM of a charged fermion
$f$. Either one or three of the DDM-photon interaction vertices
must be of EDM type. The dots indicate all other diagrams
which can be obtained from the one shown by permutation of the
interaction vertices.} \label{fig:edm}
\resizebox{4.5cm}{!}{\includegraphics{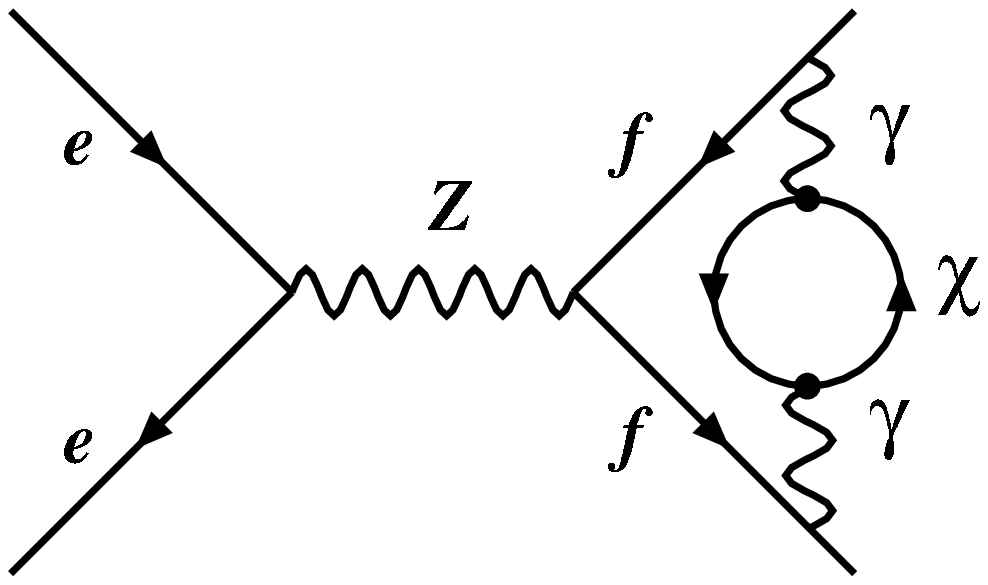}}
\caption{Lowest-order correction to $Z^0$-pole observables induced by
dipole moments of the dark-matter particle.} \label{fig:zpole}
\end{center}
\end{figure}

Furry's theorem tells us that in evaluating radiative corrections to a process
one should only keep the diagrams with an even number of photons attached to a
closed loop.\footnote{Since the theorem is valid for interactions that preserve
charge conjugation invariance we can apply it to electric and magnetic dipole
moment interactions.} Contributions with an odd number of photons attached sum
to zero. On the other hand, one must have an odd number of time-reversal-odd
(T-odd) EDM vertices in the DDM loop to generate a T-odd operator. These
considerations show that the lowest-order non-vanishing diagram must have four
photons attached to the DDM loop; diagrams with two photons attached, similar
to the one in Fig.~\ref{fig:mug2}, vanish (see above). Out of the four photons
attached to the DDM loop, either one or three can have EDM coupling to DDM.
Note that in this scenario both electric and magnetic DDM moments are necessary
to generate a dipole moment for a SM fermion. With these considerations in
mind, the lowest-order three-loop diagram that induces an EDM for a charged
fermion is shown in Fig.~\ref{fig:edm}. One obtains the following estimate for
the induced electric dipole moment: 
\begin{equation}
{\cal D}_f \sim \left[{{\cal DM}({\cal D}^2+{\cal M}^2)}\right]
{e^3 m_f^3\over (16\pi^2)^3}\ln^2{m_\chi\over m_f},
\end{equation}
where a possible double-logarithmic enhancement is included. For the electron,
the present limit is ${\cal D}_e < 4\times 10^{-27}~e$~cm, which implies
$({\cal D,\, M}) < 3 \times 10^{-13}~e$~cm for $m_\chi\sim 100$ GeV and ${\cal
D}\sim{\cal M}$. For smaller $m_\chi$ the limit becomes weaker.

There are constraints on the EDMs of other systems, such as the neutron and the
mercury atom. It is non-trivial to translate such constraints into limits on
the underlying interaction. In case of the neutron, one may attempt to treat
the neutron as a point particle for virtual-photon energies below $1~$GeV. For
higher loop momenta, photons begin seeing the quarks and the contribution to
the EDM becomes suppressed by the quark masses. In this case one may use the
above equation with $m_f \to m_n$ and no logarithmic enhancement, in order to
estimate the neutron EDM:
\begin{equation}
{\cal D}_n = \left[{{\cal DM}({\cal D}^2+{\cal M}^2)}\right]
{e^3|\kappa_n|^3 m_n^3\over (16\pi^2)^3} < 6 \times
10^{-26}~e~{\rm cm},
\end{equation}
which results in the limit $({\cal D,\,M}) \lesssim 4 \times 10^{-15}~e$~cm
(assuming ${\cal D} \simeq {\cal M}$. In the above equation, $\kappa_n=-1.91$
is the magnetic moment of the neutron. It appears because the neutron is
neutral, and couples to the photon in Fig.~\ref{fig:edm} via a magnetic-moment
interaction. The limit for the EDM of the mercury atom is much stronger than
the neutron, ${\cal D}_{Hg} \lesssim 10^{-28}~e$~cm. Unfortunately, the mercury
atom is a complicated system for which the EDM is influenced by many sources.
Therefore, we leave the Hg limit for future study.

\subsection{W Boson Mass}

The DDM can contribute to the running of the fine-structure constant for
momenta ranging up to the $Z^0$ mass. Such running will affect the relationship
between the Fermi constant $G_F$, the mass of the $W^\pm$ boson, and the
fine-structure constant at zero momentum:
\begin{equation}
m_W^2 = {\pi \alpha \over \sqrt{2} G_F}{1 \over (1 - m_W^2/m_Z^2)(1-\Delta r)},
\label{eqn:wbosonmass}
\end{equation}
where $\Delta r$ is a correction calculable in a given theory. The interaction
in Eq. (\ref{Lint}) modifies the standard expression for $\Delta r$, whereby 
$\Delta r^{New}= \Pi(m_Z^2)-\Pi(0)$. In the standard model $\Delta
r^{SM}=0.0355\pm 0.0019\pm 0.0002$. On the other hand, one can use
experimentally measured values for $\alpha$, $m_{W,Z}$, and $G_F$ in
Eq.~(\ref{eqn:wbosonmass}) to infer $\Delta r^{Exp}=0.0326\pm 0.0023$, which
gives $\Delta r^{New}<0.003$ at the 95\% confidence level. Therefore, we obtain
the limit $({\cal D}^2+{\cal M}^2)^{1/2} \lesssim 3\times 10^{-16}~e$~cm.  A
full calculation of the vacuum polarization yields the constraint shown in
Fig.~\ref{fig:concord}.  This turns out to be the strongest constraint due to
precision tests of the standard model.

\subsection{Z-Pole Observables}

The DDM will contribute to various $Z^0$-pole observables through two-loop
diagrams similar to the one shown in Fig.~\ref{fig:zpole}, at the level $\alpha
({\cal D}^2 + {\cal M}^2) m_Z^2/64 \pi^3$. Requiring that these contributions
do not exceed the $\sim 0.1\%$ precision to which $Z^0$-pole observables are
typically known \cite{PDG2002} results in the constraint $({\cal D}^2+{\cal
M}^2)^{1/2}<  10^{-14}~e$~cm. Note that in order for perturbation theory to
apply for energies ${\cal E}\sim m_Z$, one must have $({\cal D,\, M}) m_Z < 1$,
which means $({\cal D,\, M}) \lesssim 7 \times 10^{-16}~e$~cm. Interestingly,
consistency with a perturbative treatment at the $Z^0$ pole imposes much
stronger constraints on the DDM than the $Z^0$-pole observables themselves.


\begin{figure}[t]
\begin{center}
\resizebox{5cm}{!}{\includegraphics{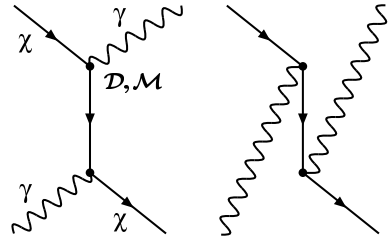}}
 \caption{The photon-DDM scattering diagram, the analog of the
Compton-scattering diagram for electric or magnetic moments.  
\label{fig:ddmphoton}}
\end{center}
\end{figure}

\subsection{Direct Production}

If kinematically allowed, DDM can be directly produced in various scattering
and decay experiments. In this case one may use the \lq\lq missing energy"
signature to constrain the DDM couplings. Here, we consider missing-energy
constraints from both low-energy ($B^+$ and $K^+$ meson decays) as well as
collider (LEP, CDF) experiments.

\subsubsection{$B^+$ and $K^+$ decays}

Searching for light ($m_\chi\lesssim 1$ GeV) dark matter using missing-energy
signatures in rare $B^+$ meson decays was originally suggested in
Ref.~\cite{Bird:2004ts}.
\begin{figure}[tb]
\begin{center}
\resizebox{8cm}{!}{\includegraphics{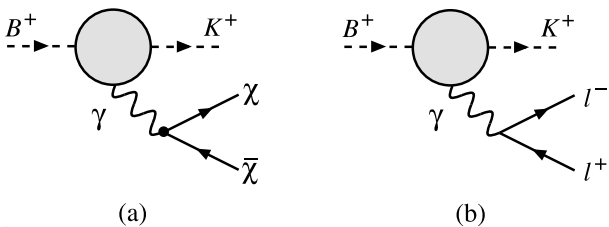}}
\caption{Photon-exchange contributions to (a) $Br\left(B^+\to
K^+\bar\chi\chi\right)$ and (b) $Br\left(B^+\to K^+ l^+l^-\right)$.
The blobs collectively represent quark flavor-changing
interactions.} \label{fig:BK-missing}
\end{center}
\end{figure}
There, data from BABAR \cite{Aubert:2003yh} and CLEO \cite{Browder:2000qr} were
used to set a limit $Br(B^+\to K^+ + {\rm invisible})\lesssim 10^{-4}$ [derived
from $Br(B^+\to K^+\bar\nu\nu$)]. This limit can be used to constrain the
dipole moments of dark matter. The diagram for $B^+\to K^+ \bar\chi\chi$ decay
is shown in Fig.~\ref{fig:BK-missing}(a). The rate for this decay can be
related to the photon-exchange contribution to $B^+\to K^+ l^+l^-|_\gamma$
shown in Fig.~\ref{fig:BK-missing}(b). Since the graphs have identical
topologies, the difference in rates will come from the difference in effective
couplings and the final-state phase-space integrals. One can estimate,
\begin{equation}
{Br\left(B^+\to
K^+\bar\chi\chi\right)\over Br\left(B^+\to K^+ l^+l^-\right)|_\gamma}
\sim \left({{\cal D}m_{B^+}\over e}\right)^2
{PS\left(K^+\bar\chi\chi\right)\over PS\left(K^+l^+l^-\right)},
\end{equation}
where $PS(\cdots)$ stand for the corresponding final-state phase-space
integrals, and $m_{B^+}=5.279$ GeV is the $B^+$ mass. Belle \cite{Abe:2001dh}
and BABAR \cite{Aubert:2001xt} find $Br(B^+\to K^+ l^+l^-)\lesssim 10^{-6}$.
Since the ratio of the phase-space integrals is of order unity, and since in
the absence of accidental cancellations $Br(B^+\to K^+ l^+l^-)|_\gamma\lesssim
Br(B^+\to K^+ l^+l^-)$, one obtains the constraint
\begin{equation}
2\times 10^{-6}\left({{\cal D}m_{B^+}\over e}\right)\lesssim 10^{-4},
\end{equation}
which leads to ${\cal D}\lesssim 3.8\times 10^{-14}~e$~cm. This constraint is
relevant for $m_\chi<(m_{B^+}-m_{K^+})/2=2.38$ GeV.

Rare $K^+$ decays can be treated in a similar manner. The relevant branching
ratios are $Br(K^+\to \pi^+ e^+e^-)=2.88^{+0.13}_{-0.13}\times 10^{-7}$ and
$Br(K^+\to \pi^+\bar\nu\nu)=0.157^{+0.175}_{-0.082}\times 10^{-9}$
\cite{PDG2002}. The resulting constraint on the dipole moment is ${\cal
D}\lesssim 1.5\times 10^{-15}~e$~cm. This constraint applies for
$m_\chi<(m_{K^+}-m_{\pi^+})/2=0.18$ GeV. We see that constraints from $B^+$ and
$K^+$ decays are not competitive with other constraints shown in
Fig.~\ref{fig:concord}.

\subsubsection{Collider experiments}

\begin{figure}[tb]
\begin{center}
\resizebox{4.5cm}{!}{\includegraphics{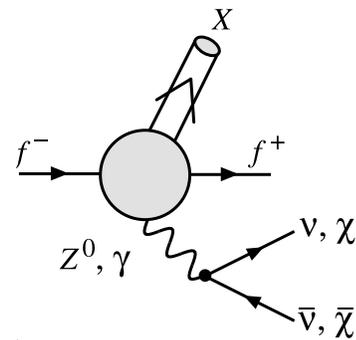}}
\caption{A typical process that would produce a missing-energy
signature in a collider experiment. Here, $X$ stands for the visible
portion of the final state. Neutrinos or DDM may carry a large
fraction of the energy but are not detected.}
\label{fig:Coll-missing}
\end{center}
\end{figure}

A typical example of a process where DDM can be directly produced in a collider
experiment is shown in Fig.~\ref{fig:Coll-missing}. Here, two fermions $f$
scatter to produce a final state containing some set of visible particles $X$
(photon, multiple jets, etc.) along with particles that are not detected. In
the SM, the latter are neutrinos.
Limits on the rate for such processes have been set by, e.g., the L3 and CDF
collaborations \cite{coll-miss}. At LEP, $X$ consisted of a single photon
whereas at CDF it consisted of one or more hadronic jets.

In order to translate constraints from collider experiments into limits on DDM
couplings one needs an analytical expression for the rate for $f\bar f\to X
\bar\chi \chi$.  Naive application of the effective Lagrangian in Eq.
(\ref{Lint}) would result in upper limits from these missing-energy searches of
roughly $10^{-17}\,e$~cm.   However, this constraint does not actually exclude
larger values of the dipole moments. Indeed, as discussed above, perturbation
theory will break down when the energy scale for the process $\cal E$ satisfies
${\cal DE}\gtrsim 1$. This means that missing-energy searches from L3 (${\cal
E} \approx 200$ GeV) and CDF (${\cal E}=1.8$ TeV) cannot be used to probe
effective dipole moments ${\cal D}>10^{-16}~e$~cm and ${\cal D}>10^{-17}~e$~cm,
respectively, unless the underlying physics that gives rise to the dipole
moment is specified. 

\subsection{Other Laboratory Constraints}

Important constraints can be obtained for millicharged particles from the Lamb
shift \cite{millichargeone,davidson} and from a targeted experiment at SLAC
\cite{slac}.  We have checked, however, that due to the different energy
dependence of the photon-dipole vertex, as opposed to the photon-millicharge
vertex, the DDM-induced correction to the Lamb shift is small for dipole
moments not eliminated by other precision measurements, such as the running of
the fine-structure constant.  Likewise, although the SLAC experiment is in
principle sensitive to neutral particles with a dipole, the production and
energy deposition of dipole particles is sufficiently small, for dipole moments
consistent with accelerator experiments, to evade detection in the SLAC
experiment.

\section{Constraints from Large-Scale Structure and the CMB}
\label{section6}

We now consider the effects of the interaction ${\cal L}_{\gamma\chi}$ on the
evolution of cosmological perturbations and their resulting imprints on the
matter power spectrum and the CMB.  A dipole moment can induce a coupling of
the dark matter to the baryon-photon fluid by scattering from photons through
the diagrams shown in Fig.~\ref{fig:ddmphoton}, or by scattering from protons,
helium nuclei, and/or electrons through the diagram shown in
Fig.~\ref{fig:dn}.  What we will show below is that the dark matter is coupled
to the baryon-photon fluid at early times, and decouples at later times.  When
the dark matter is coupled to the photon-baryon fluid, the pressure of the
plasma resists the growth of gravitational potential wells. Thus, the
short-wavelength modes of the density field that enter the horizon at early
times will have their growth suppressed relative to the standard calculation
resulting in a suppression of small-scale power.  The evolution of the
longer-wavelength modes that enter the horizon after the dark matter has
decoupled remain unaffected.  Before presenting the results of our detailed
analysis, we begin with some simple estimates.

We first show that DDM-photon scattering is negligible compared with DDM-baryon
scattering in providing  the drag force between the DDM fluid and the
baryon-photon fluid.  To do so, we first estimate the drag force per unit mass
(i.e., the deceleration) on a DDM particle that moves with a velocity $V$ with
respect to the rest frame of a blackbody at temperature $T$.  The diagrams in
Fig. \ref{fig:ddmphoton} will lead to a photon-DDM scattering cross section
$\sigma_{\chi\gamma} \sim {\cal D}^4 E_\gamma^2$.  Considering that the
momentum transfer to the DDM particle in each scatter is $\sim E_\gamma$ and
that the difference of the fluxes of photons moving in the same versus opposite
direction to the DDM particle is $\sim T^3 (V/c)$, we conclude that the
deceleration due to photon scattering is $a_\gamma \sim {\cal D}^4
T^6(V/c)/m_\chi$.

We next estimate $a_b$, the drag force per unit mass due to DDM-proton
scattering.  We first note that the peculiar velocity of the baryon-photon
fluid (obtained from the continuity equation) in the early Universe will be
$V\sim (H/k)c\delta$, where $H$ is the Hubble parameter, $k$ is the physical
wavenumber of the mode in question, and $\delta\sim10^{-5}$ is the amplitude of
the fractional density perturbation.  Since $(H/k)\lesssim1$ for modes inside
the horizon, we must have $V \lesssim 10^{-5}\,c$.  On the other hand, the
proton thermal velocity dispersion is $\bar v_p \sim (T/m_p)^{1/2}c \gtrsim
10^{-4.5}\, c$ before recombination.  Thus, for the early times of interest to
us here, the relative velocity between the DDM and the baryon-photon fluid is
small compared with the thermal proton velocities.  Thus, the appropriate
relative velocity to use in Eq. (\ref{eqn:dn}) in estimating the proton-DDM
cross section is $\bar v_p$, resulting in a DDM-proton cross section
$\sigma_{\chi b} \sim e^2 {\cal D}^2/\bar v_p^2$.  The momentum transfer per
scatter is $\sim \mu \bar v_p$, where $\mu$ is the proton-DDM reduced mass, and
the difference of the fluxes of protons moving with as opposed to against the
DDM fluid is $n_p V$, where $n_p$ is the proton density.  The drag force per
unit mass on the DDM fluid due to scattering with protons is thus $a_b \sim e^2
{\cal D}^2 (\mu/m_\chi) (V/\bar v_p) n_p$.  We also conclude from the
appearance of $\mu$ in this result that drag due to scattering from electrons
is negligible compared with baryon drag.

Since $n_p \propto T^{3}$ and $\bar v_p \propto T^{1/2}$, we find $a_b \propto
T^{2.5}$ as opposed to $a_\gamma \propto T^6$.  Thus, at early times, photon
drag dominates while baryon drag dominates at later times.  The transition
occurs at a temperature $T\sim$~GeV for values of $m_\chi$ and ${\cal D}$ of
interest to us, and such high temperatures correspond to (comoving) horizon
scales considerably smaller than the distance scales ($\gtrsim$~Mpc) probed by
large-scale structure.  We can thus neglect photon drag.  From $a_b\propto
T^{2.5}$ we infer a deceleration time for the DDM fluid $t_{\rm
dec}=V/a_b\propto T^{-2.5}$.  Since this decreases more rapidly than the Hubble
time $t_U \sim m_{\rm Pl} T^{-2}$ (where $m_{\rm Pl}\sim10^{19}$ GeV is the
Planck mass), we conclude that DDM particles are tightly coupled to the plasma
at early times and then are decoupled at later times.  With these rough
estimates, the transition temperature is $T\sim10\,{\rm keV}\,{\cal
D}_{15}^{-4}(1+m_\chi/m_p)^2$ suggesting that power on scales smaller than
$\lambda \sim10^{-2}\,{\cal D}_{15}^4(1+m_\chi/m_p)^{-2}$~Mpc will be
suppressed.  The $T^{0.5}$ dependence of the ratio of the deceleration and
expansion times suggests furthermore that the small-scale suppression will
change gradually, rather than exponentially, with wavenumber $k$.  Knowing that
the linear-theory power spectrum is measured and roughly consistent with scale
invariance down to distances $\sim$Mpc leads us to conclude that dipole moments
${\cal D} \gtrsim 5\times 10^{-15}\,(1+m_\chi/m_p)^{1/2}\,e$~cm will be ruled
out.  Strictly speaking, when $m_\chi < m_p$, the detailed calculation must
take into account the velocity dispersion of the DDM particles; our detailed
calculation below includes these effects. As seen below, the detailed analysis
leads to a slightly stronger constraint.

\subsection{Exact Equations}

The standard calculation of perturbations in an expanding universe requires
solution of the combined Einstein and Boltzmann equations for the distribution
functions of the dark matter, baryons, photons, and neutrinos including all
relevant standard-model interactions (see, e.g., Refs.
\cite{Ma:1995ey,Dodelson:2003} and references therein). Since the perturbations
are initially very small, linear perturbation theory is an excellent
approximation; this allows us to solve the perturbation equations in Fourier
space at each wavenumber $k$ independently of all other wavenumbers (modes are
uncoupled). The scattering of photons and baryons by DDM through the
interaction ${\cal L}_{\gamma\chi}$ influences the growth of cosmological
perturbations by introducing additional collision terms to the Boltzmann
equations, which ultimately result in a drag force between the DDM and the
colliding species in the equations describing the cosmological fluid (see,
e.g., Refs. \cite{Chen:2002yh,sigurdson}, which consider similar effects). 
Below we present the exact perturbation equations including the effects of dark
matter with electric or magnetic dipole moments.  Since solutions to these
equations are numerically intensive when photons and baryons are tightly
coupled through Compton scattering, we also discuss the equations appropriate
for solving for the DDM, photon, and baryon perturbations during the epoch of
tight coupling.

In the synchronous gauge the equations describing the evolution of baryons,
photons, and dark matter with an electric or magnetic dipole moment are
\begin{widetext}
\begin{eqnarray}
\dot{\delta}_{\gamma} &=& -\frac{4}{3}\theta_{\gamma}-\frac{2}{3}\dot{h}\, ,\,\,\,\,\,\,\,\,\,\,\,
\dot{\delta}_{b}  = - \theta_{b}-\frac{1}{2}\dot{h}\, ,\,\,\,\,\,\,\,\,\,\,\,
\dot{\delta}_{\chi}  = - \theta_{\chi}-\frac{1}{2}\dot{h} \, ,\cr
\dot{\theta}_{\gamma} &=& k^2 \left( \frac{1}{4}\delta_{\gamma} - 
\Sigma_{\gamma} \right) 
+ a n_{e}\sigma_{T}(\theta_{b}-\theta_{\gamma}) 
+  a n_{\chi} \<{\sigma}_{\chi \gamma}(\theta_{\chi}-\theta_{\gamma}), \cr
\dot{\theta}_{b} &=& -\frac{\dot{a}}{a}\theta_{b} 
+ c_{sb}^2k^2\delta_{b} + \frac{4\rho_{\gamma}}{3\rho_{b}} a n_{e} 
\sigma_{T} (\theta_{\gamma}-\theta_{b}) 
+ a n_{\chi}\<{\sigma v}_{\chi b}(\theta_{\chi}-\theta_{b}), \cr
\dot{\theta}_{\chi} &=& -\frac{\dot{a}}{a}\theta_{\chi} + c_{s\chi}^2k^2\delta_{\chi}
+ \frac{\rho_{b}}{\rho_{\chi }}a n_{\chi}\<{\sigma v}_{\chi
b}(\theta_{b}-\theta_{\chi})   
+ \frac{4\rho_{\gamma}}{3\rho_{\chi}}a n_{\chi}\<{\sigma}_{\chi\gamma}
(\theta_{\gamma}-\theta_{\chi}).
\label{exact_eqn}
\end{eqnarray}
\end{widetext}
While the evolution equations for the density contrast $\delta_{j} =
\delta\rho_{j}/\rho_{j}$ for each species $j \in \{\gamma,b,\chi\}$ are as in
the standard case \cite{Ma:1995ey}, as discussed above, the evolution equations
for the fluid-velocity perturbations have additional drag-force terms due to
the photon-DDM interaction.  Note that in these equations and what follows the
variable $\theta_{j} = i k V_{j}$ is the divergence of the fluid velocity in
Fourier space, $\Sigma_{j}$ is the shear, $c_{sj}$ is the intrinsic sound
speed,  and $n_{j}$ and $\rho_{j}$ are the background number and energy
densities of a particular species $j$, respectively. The variable $h$ is the
trace of the scalar metric perturbation in Fourier space (not to be confused
with the Hubble parameter), $a$ is the cosmological scale factor, and an
overdot represents a derivative with respect to the conformal time $\tau$. 
Furthermore, $\sigma_{T}$ is the Thomson cross section, while
\begin{equation}
\<{\sigma}_{\chi\gamma} = \frac{80}{21}\pi {\cal D}^{4} T_{\gamma}^2,
\end{equation}
is the appropriately thermally-averaged DDM-photon cross section, which can be
obtained from the differential cross section
\cite{Low:1954kd,Gell-Mann:1954kc},
\begin{equation}
     {d\sigma_{\chi\gamma} \over d\Omega} = {({\cal D}^4 +{\cal
     M}^4) E_\gamma^2 \over 8 \pi^2} (3-\cos^2\theta),
\end{equation}
for photon-DDM scattering.  As argued above, the photon-DDM drag term is small,
and we consider it no further in Eq. (\ref{exact_eqn}).
 
The quantity
\begin{equation}
\<{\sigma v}_{\chi b} = \frac{4(1+\xi Y)}{3\pi^2 \sqrt{\<{v_p}^2 +
\<{v_\chi}^2}} \frac{m_{\chi}}{m_{\chi} + m_{\rm p}}e^2 ({\cal
D}^2+{\cal M}^2),
\label{sigmavdb}
\end{equation}
is the appropriate thermally-averaged cross section times relative velocity for
the baryon-DDM coupling, and
\begin{align}
     \xi = 8 \frac{m_{\chi} + m_{\rm p}}{m_{\chi} + 4m_{\rm
     p}}\sqrt{\frac{\<{v_p}^2 + \<{v_\chi}^2}{\<{v_p}^2 +
     4\<{v_\chi}^2}} - 1
\label{withY}
\end{align}
is the relative efficiency for coupling to helium nuclei compared to protons. 
Appendix A provides a derivation of this collision coefficient. In these
expressions, $Y=\rho_{He}/\rho_{b} \simeq 0.24$ is the cosmological helium mass
fraction (approximating $m_{He} \simeq 4m_{p}$), $\<{v_p} = \sqrt{8 T_{b} / \pi
m_{p}}$ is the average thermal speed of the protons, $\<{v_\chi} = \sqrt{8
T_{\chi} / \pi m_{\chi}}$ is the average thermal speed of the DDM, and
$T_{\gamma}$, $T_{b}$, and $T_{\chi}$  are the photon, baryon, and DDM
temperatures respectively.  The dark-matter temperature evolves according to
\begin{eqnarray}
     \dot T_\chi &=& -2 {\dot a \over a} T_\chi  
     + {2  a \rho_b
     \<{\sigma v}_{\chi b}^{\rm T} \over m_\chi
     +m_p}(T_b-T_\chi) \cr
     &+&{8 a\rho_{\gamma} \<{\sigma}_{\chi \gamma}\over 3 m_{\chi}} (T_{\gamma} - T_{\chi}),
\end{eqnarray}
where $\<{\sigma v}_{\chi b}^{\rm T}$ is the same as the expression given in
Eq. (\ref{sigmavdb}) with the replacement of $\xi$ by $\xi^{\rm T}$ which is
given by the expression in Eq. (\ref{withY}) with the factor
$(m_\chi+m_p)/(m_\chi+4 m_p)$ replaced by $[(m_\chi+m_p)/(m_\chi+4 m_p)]^2$.
The final term, describing the dark matter heating by photons, is important at
very early times. For the dipole strength and mass range considered, the
influence is manifest only on very small length scales, below the range of
interest.

At early times, the DDM temperature $T_\chi\simeq T_b$, but at later times,
when the DDM decouples, $T_\chi$ drops relative to $T_b$.   The DDM-proton
cross section is $\propto v^{-2}$, which leads to $ \<{\sigma v}_{\chi b}
\propto \<{v_p}^{-1}$.   As a result, we cannot directly apply the results of
Ref. \cite{Chen:2002yh}, wherein a velocity-independent dark matter-baryon
interaction was assumed. However, we have verified that we recover their
results if we take a velocity-independent cross section as the source of
dark-matter--baryon drag.

\subsection{Tightly Coupled Equations}

At early times when $\tau_c^{-1} \equiv a n_e \sigma_T \gg \dot a/a$ the rapid
scattering of baryons and photons forces these species to have nearly equal
fluid velocities, and consequently the solution of the equations shown in
Eq.~(\ref{exact_eqn}) is numerically intensive. Following standard procedures
\cite{Peebles:ag,Ma:1995ey} we derive a set of equations to leading order in
the (conformal) Compton scattering time $\tau_c$ that are appropriate for
evolving the fluid variables through this epoch of tight coupling.  We first
write down an equation for the time derivative of ${\theta}_b -
{\theta}_{\gamma}$ which is usually termed the baryon-photon `slip' to leading
order in $\tau_c$,
\begin{widetext}
\begin{eqnarray}
\dot{\theta}_{b} - \dot{\theta}_{\gamma} =
\frac{2 R_{\gamma b}}{1+R_{\gamma b}}\frac{\dot{a}}{a}(\theta_{b}-\theta_{\gamma}) + 
\frac{\tau_{c}}{1+R_{\gamma b}}\left[ -\frac{\ddot{a}}{a}\theta_{b} 
+k^2\left(c_{sb}^2\dot{\delta}_b -\frac{1}{4}\dot{\delta}_{\gamma} 
- \frac{1}{2}\frac{\dot{a}}{a}\delta_{\gamma}\right)
+\frac{1}{\tau_\chi}(\dot{\theta}_\chi - \dot{\theta}_b) -
\frac{1}{2}\frac{\dot{a}}{a}\frac{1}{\tau_\chi}(\theta_\chi-\theta_b)
\right],
\end{eqnarray}
\end{widetext} 
where we have introduced the (conformal) DDM-baryon scattering time
$\tau_\chi^{-1} \equiv a n_\chi \<{\sigma v}_{\chi b}$, and $R_{\gamma b}\equiv
4 \rho_\gamma/(3\rho_b)$.  It is useful to separate this equation as a sum of
the terms not containing $\tau_\chi$ (this is just the time derivative of the
standard slip, which we denote $S_{b\gamma}$), and the new terms introduced by
the DDM coupling,
\begin{eqnarray}
\dot{\theta}_{b} - \dot{\theta}_{\gamma} = \dot{S}_{b\gamma} +
\beta \left[ (\dot{\theta}_\chi - \dot{\theta}_b) -
\frac{1}{2}\frac{\dot{a}}{a}(\theta_\chi-\theta_b) \right],
\end{eqnarray}
where 
\begin{eqnarray}
\beta = \frac{1}{1+R_{\gamma b}}\frac{\tau_c}{\tau_\chi}
\end{eqnarray}
is the parameter that controls how strongly the new interaction affects the
evolution of the slip.  In terms of these definitions the baryon-velocity
evolution equation is
\begin{widetext}
\begin{align}
\dot{\theta}_{b} = \frac{1}{1+R_{\gamma b}+\beta R_{\gamma
b}}\left\{ -\frac{\dot{a}}{a}\theta_{b} +
k^2\left[c_{sb}^2\delta_{b} + R_{\gamma
b}\left(\frac{1}{4}\delta_{\gamma}-\Sigma_{\gamma}\right)\right]
+ R_{\gamma b} \left[\dot{S}_{b \gamma}+
\beta\left(\dot{\theta}_\chi -
\frac{1}{2}\frac{\dot{a}}{a}(\theta_\chi-\theta_b)\right) \right]
\right\}.
\end{align}
\end{widetext}
The photon-evolution equation is then given by the exact expression
\begin{align}
\dot{\theta}_{\gamma} = &-\frac{1}{R_{\gamma b}} \left[
\dot{\theta}_{b} + \frac{\dot{a}}{a}\theta_{b} -
c_{sb}^2k^2\delta_{b} -
\frac{1}{\tau_\chi}(\theta_{\chi}-\theta_{b}) \right]  \\
\nonumber &+ k^2\left(\frac{1}{4}\delta_{\gamma} -
\Sigma_{\gamma}\right). 
\label{theta_g_tc}
\end{align}
We use these equations to follow the initial evolution of the baryon and photon
fluid variables and switch to the exact equations of Eq.~(\ref{exact_eqn}) at
later times. For the evolution of the DDM fluid variables we always use the
exact form of Eq.~(\ref{exact_eqn}).

\subsection{Effects on the Matter and CMB Power Spectra}

\begin{figure}
\resizebox{8cm}{!}{\includegraphics{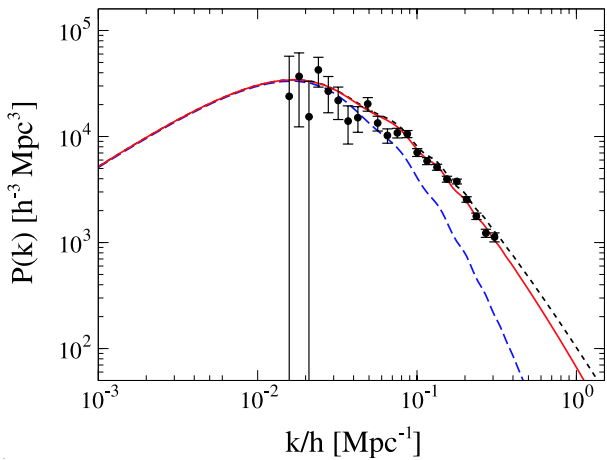}}
\caption{\label{fig:power} 
Matter power spectra including baryon-DDM drag.   The solid (red) curve is for
is for $({\cal D}^2+{\cal M}^2)^{1/2}=1.4\times10^{-15}\,e$~cm The short-dash
(black) is for $({\cal D}^2+{\cal M}^2)^{1/2}=1.0\times10^{-16}\,e$~cm The
long-dash (blue) curve is for is for $({\cal D}^2+{\cal
M}^2)^{1/2}=5\times10^{-15}\,e$~cm.  These are all for a mass of 1 GeV.  The
curves are all for the standard concordance cosmological parameters, and the
data points are from SDSS \cite{Tegmark:2003uf}}
\end{figure}

\begin{figure}
\resizebox{8cm}{!}{\includegraphics{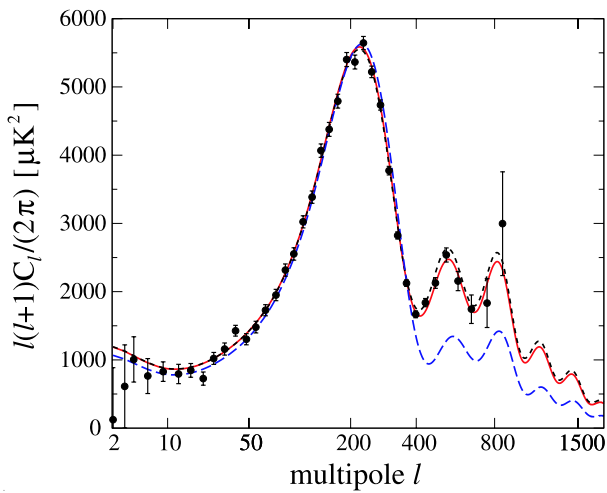}}
\caption{\label{fig:cmb} CMB power spectra including
DDM-baryon drag.  The labeling of the curves is the same as in
Fig. \protect\ref{fig:power}, and the data points are those from
WMAP \cite{Hinshaw:2003ex}}
\end{figure}

In Fig.~\ref{fig:power} we show the linear matter power spectrum and in
Fig.~\ref{fig:cmb} we show the angular power spectrum of the CMB for several
values of the dipole moment and for DDM mass $m_\chi=1$~GeV. Physically, the
effects of DDM on the matter power spectrum and CMB can be simply understood. 
Prior to matter-radiation equality the photons have a much larger density than
the baryons or the DDM and so to a first approximation completely drive the
behavior of the baryon perturbations through Compton scattering.  In turn, the
baryon perturbations drive the behavior of the DDM perturbations, very
efficiently before DDM decoupling so that the DDM density contrast
$\delta_{\chi}$ on scales that enter the horizon during this epoch track the
oscillations of the baryon-photon fluid before growing, and less efficiently
after DDM decoupling so that the baryons simply cause a drag on the growth of
$\delta_{\chi}$.  In either case the matter power spectrum is suppressed
relative to the standard case.  The behavior of the CMB angular power spectrum
can be similarly understood.  Roughly speaking,  the coupling of the DDM and
baryons increases the effective baryon loading of the plasma at early times so
that the CMB power spectra look similar to those from high-baryon models. This
is of course an imperfect correspondence as modes of larger wavelength enter
the horizon when the coupling is weaker, and so at later and later times the
evolution of the photon perturbations becomes more and more like the
standard-CDM case.  But due to geometrical projection effects modes of
wavenumber $k$ contribute to all $l \lesssim k d_{A}$ where $d_{A}$ is the
angular-diameter distance to the last-scattering surface, and so the effects of
DDM on small length scales can be noticed even on relatively large angular
scales in the CMB.

As the effect of DDM on the CMB may be partially degenerate with other
cosmological parameters, we have explored a parameter space that allows us to
constrain $m_\chi$ and $({\cal D},{\cal M})$ after marginalizing over other
cosmological parameters. We consider flat $\Lambda$CDM models and our chosen
parameter space is the dark-matter density $\Omega_m h^2$, the baryon density
$\Omega_b h^2$, the Hubble parameter $h$ in units of $100~{\rm
km~sec^{-1}~Mpc^{-1}}$, the optical depth $\tau_{\rm CMB}$ to the
last-scattering surface, and the primordial spectral index $n$.  We have
employed the Markov chain Monte Carlo technique (see, e.g.,
Ref.~\cite{Doran:2003ua}) to efficiently explore this parameter space, taking
the most recent results from SDSS \cite{Tegmark:2003uf}, WMAP
\cite{Hinshaw:2003ex}, CBI \cite{Readhead:2004gy}, VSA \cite{Dickinson:2004yr},
and SNe Ia \cite{SnIa} as our data. Note that although DDM has no effect on
observations of Type Ia supernovae, we include these data because the other
parameters we allow to vary are constrained by these observations.   We
conclude using a relative-likelihood test that cosmological measurements lead
to the bound shown in Fig. \ref{fig:concord}. The numerical calculations
confirm the qualitative behavior discussed above. Dipole moments as large as
${\cal D} \sim 10^{-17}~e~{\rm cm}$, near the upper end of our allowed
parameter space, are thus cosmologically viable.

\section{Gamma Rays}

DDM particles in the Galactic halo can annihilate to two photons through the
diagrams shown in Fig. \ref{fig:ddtoff}.  Since halo particles move with
velocities $v\simeq300$~km~sec$^{-1}\ll c$, the photons produced will be very
nearly monoenergetic with energies equal to the DDM-particle rest mass.  The
intensity at Earth of such gamma rays is obtained by integrating the
emissivity, $n_\chi^2 \<{\sigma_{\bar\chi \chi \rightarrow 2\gamma} v}$, where
$n_\chi$ is the DDM number density, along the given line of sight.  The
intensity is largest toward the Galactic center, where the dark-matter density
is largest.  In this direction, the gamma-ray intensity is then
\cite{Jungman:1995df},
\begin{equation}
     {d {\cal F} \over d\Omega} = 1.0\times10^{-10}\, {
     \sigma_{\bar\chi \chi \rightarrow 2\gamma} v \over 10^{-30}
     \, {\rm cm}^3~{\rm sec}^{-1}}m_{\rm GeV}^{-2} I\,{\rm
     cm}^{-2}~{\rm sec}^{-1}~{\rm sr}^{-1}, 
\end{equation}
where $I$ is a scaled integral of $n_\chi^2$ along a line of sight toward the
Galactic center.  The numerical coefficient is one-half that from Ref.
\cite{Jungman:1995df} since we have here particle-antiparticle annihilation
rather than Majorana annihilation.  Roughly speaking, $I\simeq3-30$ for
cored-isothermal-sphere models of the Galactic halo, while $I$ can extend up to
$\sim300$ for Navarro-Frenk-White profiles \cite{ullio}; i.e., uncertainty in
the dark-matter distribution in the inner Galaxy leads to an uncertainty of two
orders of magnitude in the predicted flux. We thus expect
\begin{eqnarray}
     {d{\cal F} \over d\Omega} &=& (3\times10^{-13} - 3\times
     10^{-11}) \cr
     & & \qquad ({\cal D}_{17}^4+{\cal
     M}_{17}^4)\, {\rm cm}^{-2}~{\rm sec}^{-1}~{\rm sr}^{-1}.  
\end{eqnarray}

To constrain dipole moments from non-observation of a gamma-ray line, we choose
to use the most conservative estimate, $I\simeq3$ for the dimensionless line
integral.  Moreover, we are not aware of any EGRET analysis that places limits
in particular to a line flux.  We thus obtain very conservative limits by using
the binned continuum fluxes for the {\it total} gamma-ray flux listed in Table
1 of Ref. \cite{ullio} and noting that a line flux in that bin cannot exceed
the measured continuum flux.  The EGRET limits apply for masses $0.1 \lesssim
(m_\chi/{\rm GeV}) \lesssim10$, and range from ${\cal D}_{17}\lesssim 180$ for
$m_\chi\lesssim$~GeV to ${\cal D}_{17}\lesssim100$ for $m_\chi\simeq10$~GeV, as
shown in Fig. \ref{fig:concord}.

Again, a few caveats are in order.  First of all, our limit is quantitatively
conservative---we chose the halo model that produces the lowest flux, and a
detailed EGRET analysis would probably yield a line-flux limit lower than what
we have assumed.  On the other hand, the strong dependence $\propto {\cal D}^4$
of the predicted flux on the dipole moment guarantees that the upper limit to
the acceptable dipole moment will not depend quite so strongly on these
details. Second, if ${\cal D}_{17}\gtrsim5$ in the mass range 100 MeV to 1 GeV,
then the correct cosmological abundance most likely requires a
particle-antiparticle asymmetry.  If so, then the annihilation rate in the halo
could be reduced far below the values we have obtained above.  We conclude by
noting that with the increased sensitivity of the Gamma-Ray Large Area Space
Telescope (GLAST), a detailed search for a line flux, and the possibility that
the actual halo model provides a more generous annihilation rate, an observable
GLAST signature may exist for masses $0.1-1$~GeV and dipole moments as low as
${\cal D}_{17}\sim10$.

\section{Discussion}
\label{section7}

In this paper we have considered the cosmology and phenomenology of dark-matter
particles with a nonzero magnetic or electric dipole moment.  We have found
that information from precision tests of the standard model, direct dark-matter
searches, gamma-ray experiments, and the CMB and large-scale structure restrict
the dipole moment to be $[{\cal D},{\cal M}]\lesssim3\times10^{-16}\,e$~cm for
masses $m_\chi\lesssim$~few GeV and  $[{\cal D},{\cal M}]\lesssim
10^{-24}\,e$~cm for larger masses.  Some of the allowed regions of parameter
space may soon be probed with GLAST and with future more sensitive
direct-detection experiments.  The electromagnetic interactions of these
particles with nuclei are coherent.  Moreover, these particles cannot
annihilate directly to neutrinos.  Therefore, searches for energetic neutrinos
from decays of the products of $\bar \chi\chi$ annihilation in the Sun or Earth
are thus likely to provide less sensitive probes than direct searches
\cite{sadoulet}.  Moreover, if there is a particle-antiparticle asymmetry, then
the energetic-neutrino flux could be reduced without altering the
direct-detection rate.

We have restricted our attention to particles with masses $m_\chi\gtrsim$~MeV,
with the notion that lower-mass particles will violate BBN limit, as discussed
toward the end of Section \ref{sec:theory}.  We also consider masses
$m_\chi\gtrsim$~MeV, as particles of lower mass will almost certainly undergo
relativistic freezeout and thus lead to unacceptable dark-matter candidates. 
However, as also noted above that if an $m_\chi\lesssim$~MeV particle has a
dipole moment $[{\cal D},{\cal M}]\lesssim 10^{-22}\,e$~cm and no other
interactions with ordinary matter, then it might still be consistent with BBN. 
Of course, such a particle will, assuming standard freezeout, have a mass
density many orders of magnitude larger than the dark-matter density.  But
suppose we were to surmise that the dark-matter density was fixed by some other
mechanism. E.g., suppose the dipole was sufficiently weak that it never came
into equilibrium.  In this case, an additional constraint to the dark-matter
dipoles can be obtained from energy-loss arguments applied to stars in globular
clusters.  Such arguments eliminate dipole moments $[{\cal D},{\cal M}]
\lesssim 6\times10^{-23}\,e$~cm for masses $m_\chi\lesssim5$~keV
\cite{raffelt}.  We have also considered constraints from astrophysical
phenomena such as the stability of the Galactic disk, lifetime of compact
objects, and annihilations in the solar neighborhood \cite{Starkman:1990nj},
and find that these constraints on the mass and interaction strengths are not
competitive with those presented here.

It would be of interest to attempt to embed this scenario in a consistent
particle physics model. We might find links between baryonic and non-baryonic
matter abundances, the dark matter electric dipole moment and the CP violation
needed for baryogenesis, and the magnetic moments of dark matter and baryons.
However, such model building is beyond the scope of the present study. Our
approach throughout has been entirely phenomenological, as we have been
motivated by the desire to answer the question, ``How dark is `dark'?"

\begin{acknowledgments}

We thank S. Golwala, J. Albert, and M. Zaldarriaga for useful discussion. KS
acknowledges the support of a Canadian NSERC Postgraduate Scholarship.  This
work was supported in part by NASA NAG5-9821 and DoE DE-FG03-92-ER40701 (at
Caltech) and NSF PHY-0099543 (at Dartmouth). MD and RC thank Caltech for
hospitality during the course of this investigation.

\end{acknowledgments}

\appendix
\section{Derivation of the Baryon-Dark Matter Collision Term}

To determine how the cosmological perturbation equations for baryons and dark
matter are altered when we bestow the dark matter with a magnetic or electric
dipole moment, we must formally evaluate the collision operator of the
general-relativistic Boltzmann equation in a given gauge
\cite{KolbTurnerbook,Dodelson:2003,Ma:1995ey} for the dipole interaction of Eq.
(\ref{Lint}).  We have completed this calculation in detail, and find that the
dipole interaction produces a drag force proportional to the relative velocity
$V=v_\chi-v_b$ of the dark-matter and baryon fluids.  As the relative velocity
is gauge invariant in linear perturbation theory and all scatterings are local
processes, we may thus take a simpler, more physically transparent approach and
just evaluate this drag force using nonrelativistic statistical mechanics.  It
is this approach we now present.

We wish to calculate the drag force per unit mass, or deceleration, due to
collisions with protons to the dark-matter fluid as it passes through the
baryon-photon fluid. Comoving scales $\lambda\gtrsim$~Mpc enter the horizon
when the cosmological temperature is $T\lesssim10$ eV, when the DDM particles
(which are restricted to $m_\chi\gtrsim$~MeV) are nonrelativistic.  We may thus
consider thermal velocity distributions for nonrelativistic baryons and dark
matter. Since the drag force can only depend on the dark-matter--baryon
relative velocity, we take the baryon fluid to be at rest and the dark-matter
fluid to have a velocity of magnitude $V$ in the $\hat x$ direction.  Then, the
proton phase-space distribution is
\begin{equation}
     f_p(\vec v_p) = {n_p \over (2\pi)^{3/2} \bar v_p^3} e^{-v_p^2
     /2\bar v_p^2},
\end{equation}
where $\bar v_p=(kT_p/m_p)^{1/2}$ is the proton velocity
dispersion and $n_p$ the proton number density, and
\begin{equation}
     f_\chi(\vec v_\chi) = {n_\chi \over (2\pi)^{3/2} \bar v_\chi^3}
     \exp\left[- {(\vec v_\chi -V \hat x)^2 \over 2\bar v_\chi^2}\right],
\end{equation}
is the dark-matter phase-space distribution, with $\bar v_\chi =( k
T/m_\chi)^{1/2}$.  Recall also that we expect $V \ll \bar v_p$, as
discussed above.

The drag force per unit mass is obtained by integrating the momentum transfer
per collision over all collisions between protons and dark-matter particles. 
From the symmetry of the problem, the deceleration of the dark-matter fluid
will be in the $\hat x$ direction, and it will have a magnitude,
\begin{eqnarray}
     a_x = & &{1\over n_\chi} \int d^3 v_\chi f_\chi(\vec v_\chi) \int d^3 v_p
     f_p(\vec v_p) |\vec v_p -\vec v_\chi|\cr
     & & \qquad \times \int d\Omega {d\sigma
     \over d\Omega} (v_{\chi xf}-v_{\chi xi}).
\label{eqn:ax}
\end{eqnarray}
Here $\Omega=(\theta,\phi)$ is the scattering angle in the center-of-mass
frame, and $v_{\chi xf}-v_{\chi xi}$ is the difference between the final and
initial $x$ component of the dark-matter--particle velocity; the difference is
the same in the center-of-mass and laboratory frames.  The differential cross
section $d\sigma/d\Omega$ is that given in Eq. (\ref{eqn:dn}).

Consider an individual scattering event.  Let $\alpha$ be the angle that $\vec
v_p -\vec v_\chi$ makes with the $\hat x$ direction; this is then the angle
that $\vec v_\chi$ makes with the $\hat x$ axis in the center-of-mass frame,
and the magnitude of the initial and final dark-matter velocities in the
center-of-mass frame is $v_\chi^{\rm cm}=m_p v /(m_p+m_\chi)$, where $v\equiv
|\vec v_p-\vec v_\chi|$ is the relative velocity.  The initial $\hat x$
component of the dark-matter velocity in the center-of-mass frame is then
$v_{\chi xi}=v_\chi^{\rm cm} \cos\alpha$. The scattering angles $\theta$ and
$\phi$ are then the polar and azimuthal angles that the scattered dark-matter
velocity make with the initial velocity in the center-of-mass frame.  By
rotating this coordinate system by an angle $\alpha$ about the $\hat z$ axis to
align it with the laboratory $\hat x$ axis, we find $v_{\chi xf}=v_\chi^{\rm
cm}(\cos\alpha\cos\theta - \sin\alpha\sin\theta\sin\phi)$.  Thus,
\begin{equation}
     \int d\Omega {d\sigma \over d\Omega}(v_{\chi xf}-v_{\chi xi}) =
     -{m_p Z^2 e^2 {\cal D}^2 \cos\alpha \over 2 \pi (m_p+m_\chi) v}.
\end{equation}
Completing the integral in Eq. (\ref{eqn:ax}) in the limit $V
\ll (\bar v_p,\bar v_\chi)$, we find
\begin{equation}
     a_x = {2 \over 3 \pi (2\pi)^{1/2}}{(Ze{\cal D})^2 m_p n_p \over
     m_p + m_\chi} {V \over \sqrt{\bar v_p^2 + \bar v_\chi^2}}.
\end{equation}
Taking into account the definition $\<{v_p}= \sqrt{8 T_p/(\pi m_p)}$, this drag
force leads to the drag-force term in Eqs. (\ref{exact_eqn}), (\ref{sigmavdb}),
and (\ref{withY}), when including the simple corrections for a mass fraction
$Y$ of helium.





\begin{thebibliography}{99}

\bibitem{Zwicky:1937}
F.~Zwicky,
Astrophys.\ J. {\bf 86}, 217 (1937).

\bibitem{Tremaine:we}
S.~Tremaine and J.~E.~Gunn,
Phys.\ Rev.\ Lett.\  {\bf 42}, 407 (1979).

\bibitem{pdg} See, e.g., {\tt http://pdg.lbl.gov}.

\bibitem{Jungman:1995df}
G.~Jungman, M.~Kamionkowski, and K.~Griest,
Phys.\ Rept.\  {\bf 267}, 195 (1996).

\bibitem{axionreviews} M. S. Turner, Phys. Rept. {\bf 197}, 67
     (1990); G. Raffelt, Phys. Rept. {\bf 198}, 1 (1990); K. van
     Bibber and L. Rosenberg, {\bf 325}, 1 (2000).

\bibitem{Starkman:1990nj}
G.~D.~Starkman, A.~Gould, R.~Esmailzadeh, and S.~Dimopoulos,
Phys.\ Rev.\ D {\bf 41}, 3594 (1990).

\bibitem{Spergel:1999mh}
D.~N.~Spergel and P.~J.~Steinhardt,
Phys.\ Rev.\ Lett.\  {\bf 84}, 3760 (2000).

\bibitem{Gould:gw}
A.~Gould, B.~T.~Draine, R.~W.~Romani, and S.~Nussinov,
Phys.\ Lett.\ B {\bf 238}, 337 (1990).

\bibitem{millichargeone} See, e.g., S. Davidson, S. Hannestad, and
     G. Raffelt, JHEP 05, 003 (2000).

\bibitem{millichargetwo} S. L. Dubovsky,
     D. S. Gorbunov, and G. I. Rubtsov, JETP Lett. 79, 1
     (2004) [ Pisma Zh.Eksp.Teor.Fiz. {\bf 79}, 3 (2004)].

\bibitem{Fermi:1947}
E.~Fermi and E.~Teller,
Phys. Rev. {\bf 72}, 399 (1947). 

\bibitem{bbn} S. Burles et al., Phys. Rev. Lett. {\bf 82}, 4176
     (1999).

\bibitem{KolbTurnerbook} E. W. Kolb and M. S. Turner, {\it The
     Early Universe} (Addison-Wesley, Redwood City, 1990).

\bibitem{CMBconstraints}
        P. de Bernardis. {\it et al.}, 
        Nature {\bf 404}, 955 (2000);  S. Hanany {\it et al.},
        Astrophys. J. Lett. {\bf 545}, L5 (2000);
        N. W. Halverson {\it et al.}, 
         Astrophys. J. 568, 38 (2002); B. S. Mason {\it et al.},
         Astrophys. J. {\bf 591}, 540 (2003); A. Benoit {\it et al.},
         Astron. Astrophys. {\bf 399}, L2 (2003).  D.~N.~Spergel {\it et al.},
         Astrophys.\ J.\ Suppl.\  {\bf 148}, 175 (2003).
 
\bibitem{Griest:1989wd}
K.~Griest and M.~Kamionkowski,
Phys.\ Rev.\ Lett.\  {\bf 64}, 615 (1990).

\bibitem{KamTur90} 
M. Kamionkowski and M. S. Turner,
Phys.\ Rev.\ D {\bf 42}, 3310 (1990).

\bibitem{Foldy:1951}
L.~L.~Foldy,
Phys.\ Rev.\ {\bf 83}, 688 (1951).

\bibitem{CDMS04}
D.~S.~Akerib {\it et al.}  [CDMS Collaboration],
arXiv:astro-ph/0405033.

\bibitem{Akerib:2003px}
D.~S.~Akerib {\it et al.}  [CDMS Collaboration],
Phys.\ Rev.\ D {\bf 68}, 082002 (2003).

\bibitem{cresst} 
G. Angloher {\it et al.},
Astropart.\ Phys.\ {\bf 18}, 43 (2002).

\bibitem{McGuire:2001qj}
P.~C.~McGuire and P.~J.~Steinhardt,
arXiv:astro-ph/0105567.

\bibitem{Rich:1987st}
J.~Rich, R.~Rocchia, and M.~Spiro,
Phys.\ Lett.\ B {\bf 194}, 173 (1987).

\bibitem{McCammon:2002gb}
D.~McCammon {\it et al.},
Astrophys.\ J.\  {\bf 576}, 188 (2002).

\bibitem{Bennett:2002jb}
G.~W.~Bennett {\it et al.}  [Muon g-2 Collaboration],
Phys.\ Rev.\ Lett.\  {\bf 89}, 101804 (2002)
[Erratum-ibid.\  {\bf 89}, 129903 (2002)].
 
\bibitem{PDG2002}
K.~Hagiwara {\it et al.} [Particle Data Group], 
Phys.\ Rev.\  D {\bf 66}, 010001 (2002). 

\bibitem{Bird:2004ts}
C.~Bird, P.~Jackson, R.~Kowalewski, and M.~Pospelov,
arXiv:hep-ph/0401195.

\bibitem{Aubert:2003yh}
B.~Aubert {\it et al.}  [BABAR Collaboration],
arXiv:hep-ex/0304020.

\bibitem{Browder:2000qr}
T.~E.~Browder {\it et al.}  [CLEO Collaboration],
Phys.\ Rev.\ Lett.\  {\bf 86}, 2950 (2001) [arXiv:hep-ex/0007057].

\bibitem{Abe:2001dh}
K.~Abe {\it et al.}  [BELLE Collaboration],
Phys.\ Rev.\ Lett.\  {\bf 88}, 021801 (2002)
[arXiv:hep-ex/0109026].

\bibitem{Aubert:2001xt}
B.~Aubert {\it et al.}  [BABAR Collaboration],
arXiv:hep-ex/0107026.

\bibitem{coll-miss}
P.~Achard {\it et al.}  [L3 Collaboration],
Phys.\ Lett.\ B {\bf 587}, 16 (2004); D.~Acosta  [CDF
Collaboration],
Phys.\ Rev.\ Lett.\  {\bf 92}, 121802 (2004); T.~Affolder {\it et
al.}  [CDF Collaboration],
Phys.\ Rev.\ Lett.\  {\bf 88}, 041801 (2002).

\bibitem{davidson} S. Davidson, B. Campbell, and D. Bailey,
     Phys. Rev. D {\bf 43}, 2314 (1991)

\bibitem{slac} A. A. Prinz {\it et al.}, Phys. Rev. Lett. {\bf
     81}, 1175 (1998).

\bibitem{Ma:1995ey}
C.~P.~Ma and E.~Bertschinger,
Astrophys.\ J.\  {\bf 455}, 7 (1995).

\bibitem{Dodelson:2003}
Scott Dodelson,
{\it Modern Cosmology} (Academic Press, Amsterdam, 2003).

\bibitem{Chen:2002yh}
X.~L.~Chen, S.~Hannestad, and R.~J.~Scherrer,
Phys.\ Rev.\ D {\bf 65}, 123515 (2002).

\bibitem{sigurdson} K. Sigurdson and M. Kamionkowski,
     Phys. Rev. Lett. {\bf 92}, 171302 (2004).

\bibitem{Low:1954kd}
F.~E.~Low,
Phys.\ Rev.\  {\bf 96}, 1428 (1954).

\bibitem{Gell-Mann:1954kc}
M.~Gell-Mann and M.~L.~Goldberger,
Phys.\ Rev.\  {\bf 96}, 1433 (1954).

\bibitem{Peebles:ag}
P.~J.~E.~Peebles and J.~T.~Yu,
Astrophys.\ J.\  {\bf 162}, 815 (1970).

\bibitem{Doran:2003ua}
M.~Doran and C.~M.~Mueller,
arXiv:astro-ph/0311311.

\bibitem{Tegmark:2003uf}
M.~Tegmark {\it et al.}  [SDSS Collaboration],
Astrophys.\ J.\  {\bf 606}, 702 (2004).

\bibitem{Hinshaw:2003ex}
G.~Hinshaw {\it et al.},
Astrophys.\ J.\ Suppl.\  {\bf 148}, 135 (2003).

\bibitem{Readhead:2004gy}
A.~C.~S.~Readhead {\it et al.},
arXiv:astro-ph/0402359.

\bibitem{Dickinson:2004yr}
C.~Dickinson {\it et al.},
arXiv:astro-ph/0402498.

\bibitem{SnIa} A.~G.~Riess {\it et al.},
arXiv:astro-ph/0402512;
J.~L.~Tonry {\it et al.},
Astrophys.\ J.\  {\bf 594}, 1 (2003);
R.~A.~Knop {\it et al.},
arXiv:astro-ph/0309368.
B.~J.~Barris {\it et al.,} Astrophys. J. {\bf 602}, 571 (2004).

\bibitem{ullio} L. Bergstrom, P. Ullio, and J. Buckley,
     Astropart. Phys. {\bf 9}, 137 (1998).

\bibitem{sadoulet} M. Kamionkowski {\it et al.}, Phys. Rev. Lett. {\bf
     74}, 5174 (1995).

\bibitem{raffelt} G. Raffelt, {\it Stars as Laboratories for
     Fundamental Physics} (University of Chicago Press, Chicago,
     1996).


\end{thebibliography}
\end{document}